%% file: Non_minimal_paper.tex
\documentclass[11pt, oneside]{article} 
\pdfoutput=1  	
\usepackage[english]{babel}
\usepackage{geometry}                		
\usepackage{graphicx}				
\usepackage{amssymb}
\usepackage{amsmath}
\usepackage{tikz}
\usepackage{placeins}
\usepackage{enumitem}
\usepackage{commath}
\usepackage{braket}
\usepackage{subcaption}
\usepackage{float}
\usepackage{ulem}

\usepackage{hyperref}
\usepackage{xcolor}
\hypersetup{
  colorlinks   = true, 
  urlcolor     = red, 
  linkcolor    = blue, 
  citecolor   = blue 
}

\numberwithin{equation}{section}

\usepackage{csquotes}
\usepackage[
    style=numeric-comp,
    bibstyle=phys,
    sorting=none,
    maxnames = 5,
    biblabel=brackets,
    backend=biber,
    eprint=true
]{biblatex}
\newcommand{\gpi}{\pi}
\newcommand{\me}{\mathrm{e}}
\DeclareRobustCommand{\orderof}{\ensuremath{\mathcal{O}}}

\DeclareRobustCommand{\d}{\ensuremath{\mathrm{d}}}
\DeclareRobustCommand{\half}{\ensuremath{\frac{1}{2}}}
\DeclareRobustCommand{\mpl}{\ensuremath{M_{\mathrm{Pl}}}}

\newcommand{\zeroitem}{\refstepcounter{enumi}\item[(0)]}

\title{{\LARGE{\textbf{Lightish but clumpy:\\
scalar dark matter from inflationary fluctuations}}}}
\author{Gonzalo Alonso\,-{\'A\!}lvarez and Joerg Jaeckel \vspace{3mm} \\  \textit{ \normalsize Institut f{\"u}r Theoretische Physik, Universit{\"a}t Heidelberg,}\\ \textit{\normalsize Philosophenweg 16, 69120 Heidelberg, Germany}}
\date{}							

\begin{document}

\maketitle

\begin{abstract}
It has recently been shown~\autocite{graham_vector_2016} that light vector particles produced from inflationary fluctuations can give rise to the dark matter in the Universe. A similar mechanism has been discussed in~\autocite{cosme_scale-invariant_2018} for a non-minimally coupled scalar enjoying a Higgs portal interaction.
We discuss in detail how such a generation of dark matter works in a minimal setup of a massive scalar non-minimally coupled to gravity.  For suitable values of the non-minimal coupling any initial constant value of the field is suppressed during inflation. At the same time, the quantum fluctuations acquired during inflation give rise to a peaked energy density power spectrum. Isocurvature constraints can be avoided since nearly all of the energy is concentrated in fluctuations too small to be observed in the CMB. For masses \(\gtrsim {\rm eV}\) and sufficiently high inflation scale the energy contained in these fluctuations is sufficient to account for the cold dark matter of the Universe. At small scales \(\ell_{\rm today}\sim 10^{4}\,{\rm km}\sqrt{m/{\rm eV}}\sim 10^{-4}\,{\rm AU}\sqrt{m/{\rm eV}}\)  fluctuations are large and we therefore expect a rather clumpy nature of this form of dark matter.
\end{abstract}

\newpage

\input{Introduction}

\input{Dark_matter}

\input{Quantum_computation}

\input{Conclusions}

\section*{Acknowledgements}
The authors would like to thank J.~Berges, A.~Chatrchyan, J.~Rubio and G.~Ballesteros for interesting discussions.
This project has received funding from the European Union's Horizon 2020 research and innovation programme under the Marie Sklodowska-Curie grant agreement No $674896$ (ITN ELUSIVES).

\appendix
\input{Jordan_Einstein_frames}

\input{Power_spectra}

\printbibliography

\end{document}

%% file: Introduction.tex

\section{Introduction}\label{sec:introduction}
Bosonic fields are interesting and well motivated candidates for the dark matter in the Universe.
Both new heavy particles as well as fields in the intermediate and low mass ranges are phenomenologically viable possibilities.

Heavy new degrees of freedom in the GeV and TeV mass range have traditionally been very popular, as a relic density compatible with the observed dark matter abundance may in a natural way be thermally produced through a freeze-out process \autocite{lee_cosmological_1977,ellis_supersymmetric_1984,goldberg_erratum:_2009,kolb_early_1990}.
Direct detection experiments, indirect observations and limits coming from colliders (see~\autocite{undagoitia_dark_2016},  \autocite{slatyer_tasi_2017} and \autocite{kahlhoefer_review_2017}, respectively, for recent reviews on these areas) are able to strongly constrain this paradigm.
It is therefore timely to study alternative scenarios that could generate dark matter in this mass range and, perhaps even more importantly, to explore different mass regimes.

Prominent examples of light bosons as dark matter candidates are axions, axion-like particles (ALPs) and hidden photons~\autocite{preskill_cosmology_1983,abbott_cosmological_1983,dine_not-so-harmless_1983,arias_wispy_2012,jaeckel_family_2014,ringwald_exploring_2012,marsh_axion_2016}.
A significant number of direct detection experiments are currently underway, while novel ideas and setups are being developed at an astonishing rate (see~\autocite{irastorza_new_2018} for a recent review).
Until recently, most efforts were concentrated on relatively low masses where the coherence and de Broglie wavelengths of the bosonic fields are macroscopic. 
However, new ideas for the detection of ${\rm meV}-{\rm keV}$ mass particles via absorption in superconductors, semiconductors as well as other materials are being brought forward~\autocite{hochberg_detecting_2016,hochberg_absorption_2017} (see also~\autocite{battaglieri_us_2017} and references therein). 
This nicely complements absorption signals for masses $\gtrsim 10$~eV which can be searched for in large scale WIMP dark matter detectors~\autocite{the_xenon100_collaboration_first_2014,xenon_collaboration_search_2017,akerib_first_2017,bloch_searching_2017} or neutrinoless double beta decay experiments~\autocite{abgrall_new_2017}.

While the observational situation is not yet fully clear, the $3.5$~keV line provides an interesting hint for a dark matter particle in the keV mass range~\autocite{bulbul_detection_2014,boyarsky_unidentified_2014}. 
Possible explanations include sterile neutrinos~\autocite{bulbul_detection_2014,boyarsky_unidentified_2014} but interestingly also scalars \autocite{cosme_scalar_2018} and axion-like particles~\autocite{jaeckel_3.55_2014}.  

The standard production mechanism for light bosons in the early Universe is the misalignment mechanism~\autocite{preskill_cosmology_1983,abbott_cosmological_1983,dine_not-so-harmless_1983,arias_wispy_2012,jaeckel_family_2014}. 
For the paradigmatic example of pseudo-Goldstone bosons, the combination of a weak coupling (i.e. high symmetry breaking scale) and masses $\gtrsim 1\,{\rm eV}$ requires a significant amount of fine tuning to avoid producing an overabundance of dark matter~\autocite{jaeckel_3.55_2014}\footnote{For a recent modification of the minimal pseudo-Goldstone boson setup, breaking the usual relation between the naturally produced dark matter abundance and the coupling to other fields see~\autocite{alonso-alvarez_exploring_2017}.}. 
At the same time, thermal production is in conflict with structure formation for masses $\lesssim\,{\rm keV}$ and may also be problematic for larger masses if interactions with the Standard Model particles are very weak. 
This motivates the search for new production mechanisms\footnote{An interesting mechanism for a potentially less tuned production of very light $\lesssim \mu\,{\rm eV}$ bosons has recently been discussed in~\cite{graham_stochastic_2018,guth_qcd_2018}.}. 

For vector particles, it has recently been shown in~\cite{graham_vector_2016} that a sufficient abundance of dark matter can be produced from the quantum fluctuations of the vector field present during inflation for masses $\gtrsim 10^{-5}$~eV.
The main goal of the present paper is to show that scalar and pseudoscalar particles non-minimally coupled to gravity are subject to a similar production mechanism and can provide the correct dark matter abundance for masses in the range $\gtrsim$~eV. The scenario is also viable for much larger masses depending on the inflationary scale, but for very heavy fields other production mechanisms \autocite{ford_gravitational_1987, chung_superheavy_1998, chung_gravitational_2001, ema_gravitational_2015, ema_gravitational_2016, markkanen_dark_2017,ema_production_2018, garny_planckian_2016, tang_pure_2016, tang_thermal_2017, garny_theory_2018} might dominate over the one presented in this work.
The inflationary production of scalar dark matter with intermediate masses has been studied by various authors~\autocite{nurmi_inflationary_2015,kainulainen_isocurvature_2016,bertolami_scalar_2016,cosme_scale-invariant_2018}, focusing on the ``Higgs portal" scenario. We provide a more minimal setup where neither self-interactions nor direct couplings to Standard Model fields are assumed.  This already leads to a different cosmological evolution and a modified range of masses and inflation scales that yield the correct dark matter abundance. Moreover, we carefully follow the evolution of the power spectrum of fluctuations and consider the particularities of the behaviour of the higher momentum modes. We find significant qualitative and quantitative differences between our formalism and a treatment of the field as quasi-homogeneous.

\bigskip

It is a well known fact that light fields present during inflation can be excited coherently~\autocite{linde_particle_2005, birrell_quantum_1982, mukhanov_introduction_2007}.
This phenomenon can be understood as arising from the fact that an observer in de Sitter space will feel a thermal bath at the Hawking temperature \(T_\mathrm{H}=H/2\pi\)~\autocite{gibbons_cosmological_1977}.
Small scale modes acquire quantum fluctuations which are stretched and grow in amplitude as the Universe expands.
Eventually, these fluctuations transition into a classical regime when they leave the horizon.

If such field excitations survive the evolution until present time, they can contribute to the energy budget of the Universe.
In particular, if we are dealing with a sufficiently weakly interacting particle that is cosmologically stable, it is possible that these particles make up the dark matter that we observe today. 
For vectors this has been realized in~\autocite{graham_vector_2016}.
Here, we will show that such a mechanism is also viable for scalars.

The main issue with this scenario has to do with the generation of isocurvature perturbations.
In general, the quantum fluctuations induced in the inflaton field are uncorrelated with the ones acquired by other light fields.
In the cosmological standard model, the inflaton is responsible for the generation of the curvature (or \textit{adiabatic}) perturbations whose spectrum perfectly matches the one imprinted in the cosmic microwave background (CMB).
As the inflaton decays during reheating, these fluctuations are passed on to the different fluids that populate the Universe.
However, if dark matter is not produced from the decay of the inflation but rather is already present during inflation, its fluctuations will not match the ones that the rest of the fluids have: they are isocurvature rather than adiabatic.
This mismatch in the spectrum of fluctuations would leave an imprint in the CMB which has not been observed by the Planck mission \autocite{planck_collaboration_planck_2018}.

The consequence is that the amplitude of isocurvature fluctuations in the dark matter field has to be very small compared to the adiabatic one.
There is however a caveat: this statement only holds for the very large scales that are observed by the Planck satellite.
It is perfectly allowed by observations that the dark matter field enjoys isocurvature fluctuations as long as they are ``hidden" at low scale modes.
This is possible, for instance, if the power spectrum of the dark matter field is peaked at some intermediate scale much below the ones probed by CMB observables.
Such a possibility was studied in \autocite{graham_vector_2016} in the context of a massive vector field.
In this paper we see that this is also the case for a scalar field non-minimally coupled to gravity. 
This is in line with the findings of~\autocite{bertolami_scalar_2016,cosme_scale-invariant_2018}, who discussed a suppression of the isocurvature perturbations of a scalar field which during inflation acquires a large mass from couplings to other fields, such as the inflaton, and a non-minimal coupling, respectively.
In this work, we investigate the cosmological evolution in more detail and keep track of the wavelength-dependent evolution of fluctuations during the different eras. 
In short, the very long-range fluctuations accessible in CMB measurements are suppressed during inflation because of the presence of a $\mathcal{O}(H_I)$ effective mass arising from the non-minimal coupling. During radiation domination the Ricci scalar drops to $R\simeq 0$, stopping the suppression of the superhorizon modes and thus avoiding a further dilution of the energy density stored in shorter wavelength modes. The result at late times is a peaked power spectrum with order one fluctuations on very small scales.

\bigskip

Throughout this work, we only deal with gravitational interactions of a non-minimally coupled scalar field, without assuming the existence of any further couplings to the inflaton or to standard model fields.
This is motivated by the fact that the only current evidence for the existence of dark matter comes from the gravitational sector. 
Indeed, the misalignment mechanism that is usually employed to produce low mass bosonic fields also does not rely on any interactions with other particles.
Further scenarios where dark matter is produced gravitationally and where it only interacts through the metric have been considered in the past.
So far, studies have focused on the production at the end of inflation or during reheating \autocite{ford_gravitational_1987, chung_superheavy_1998, chung_gravitational_2001, ema_gravitational_2015, ema_gravitational_2016,markkanen_dark_2017, ema_production_2018}, and were concerned primarily with very heavy scalar fields.
The possibility that dark matter was produced thermally through graviton-mediated processes has been considered in \autocite{ema_production_2018, garny_planckian_2016, tang_pure_2016, tang_thermal_2017, garny_theory_2018}, again for heavy scalar fields with masses above \(\sim 1\) TeV.
Our approach is essentially different, as we focus on the production purely due to the expansion during inflation and deal with lighter masses.

We would like to stress that similar to the misalignment mechanism, the production mechanism considered here is independent of additional (weak) interactions of the field with Standard Model particles. 
The resulting dark matter can therefore be equipped with a wide variety of very weak interactions with the Standard Model, e.g. two-photon couplings, derivative couplings to fermions and many others, opening a wide range of phenomenological possibilities (see~\autocite{jaeckel_low-energy_2010,graham_new_2013,jaeckel_family_2014,irastorza_new_2018} for some examples).
Another natural example would be the dimension $4$ coupling to the Higgs field, usually known as the Higgs portal coupling, whose implications for the inflationary or misalignment produced dark matter have been studied in \autocite{cosme_scale-invariant_2018}\footnote{Higgs portal dark matter from thermal production~\autocite{silveira_scalar_1985,burgess_minimal_2001,bento_cosmological_2001} is an entire field on its own.}.

Moreover, we note that this mechanism is generic and not necessarily only relegated to dark matter production\footnote{See~\autocite{bettoni_quintessential_2018} for a construction where baryogenesis is achieved via a non-minimally coupled scalar.}.
Indeed, any scalar field with a non-minimal coupling to gravity present during inflation is bound to acquire the spectrum of fluctuations predicted in this work, even the ones that are not stable on cosmological time scales.
This may have interesting implications in cosmology when applied, for instance, to the Higgs boson.
In what follows, we focus on the generation of dark matter and leave the study of this possibility for future work.

The paper is structured as follows.
In \secref{sec:dark_matter} we present the essential phenomenological results. We discuss the relevant features of the evolution of non-minimally coupled scalar fields as well as their fluctuations (sketched in Fig.~\ref{fig:master_plot})). From this we compute the power spectrum (cf. Fig.~\ref{fig:field_power_spectrum}) and the relic abundance and see how isocurvature perturbations can be suppressed at large scales resulting in the viable parameter space for the scalar fields to be the dark matter of the Universe (shown in Fig.~\ref{fig:dark_matter}).
In \secref{sec:quantum_computation}, we give a more in depth discussion of the generation of the fluctuations based on methods of quantum field theory in curved spacetimes.
Finally, \secref{sec:conclusions} is devoted to the conclusions and discussion about the potential phenomenological implications of this scenario.

%% file: Dark_matter.tex

\section{Generating dark matter from inflationary fluctuations}\label{sec:dark_matter}
In this section we discuss the main features of the generation of dark matter from inflationary fluctuations of non-minimally coupled scalar fields.

After introducing the model we first study the evolution of the homogeneous field. 
For a suitable range of the non-minimal coupling the homogeneous field is completely suppressed, ensuring independence of the results from the initial conditions. 

We then turn to the evolution of the fluctuations that give rise to the dark matter. 
We develop the power spectrum and the energy density it contains and discuss the features that allow to avoid isocurvature constraints. 

During this whole section we will treat the evolution of the field with classical equations of motion. 
As the fluctuations arise from the quantum fluctuations of the field during inflation that turn classical only later on, one may wonder if this is correct. 
The upshot is that as long as the evolution is linear, the mode functions of the creation and destruction operators follow the classical equations of motion. 
For appropriate quantities such as the power spectrum, the evolution can then be directly extracted from these mode functions and hence from the classical evolution. 
We provide more details on this in \secref{sec:quantum_computation}.

\subsection{Model setup}
Our starting point is the action of a scalar field \( \phi \) non-minimally coupled to gravity in the following way
\begin{equation}\label{eq:action}
S = \int \mathop{\d^4x} \sqrt{-g} \left( \half \left( M^2 - \xi \phi^2\right) R - \half \partial_\mu \phi \partial^\mu \phi - \half m^2\phi^2 + \mathcal{L}_{\mathrm{back}}\left( \sigma_i \right) \right).
\end{equation}
Here, \(M\) corresponds to the Planck mass in the non-minimally coupled frame and \(\mathcal{L}_{\mathrm{back}}\left( \sigma_i \right)\) refers to the action of any other field(s) \(\sigma_i\) present in the Universe. At early times they may dominate its energy density. For simplicity we take them into account by specifying a FRW background with the metric \(\d s^2 = -\d t^2 + a^2(t) \d \mathbf{x}^2\) ($t$ being the physical time) and setting the Hubble scale to an appropriate value.
The normalisation of the coupling constant \(\xi\) is chosen such that \(\xi=1/6\) corresponds to the case of a conformal coupling.
For the purposes of the present paper we will focus on $\xi\geq 0$.

Due to the non-minimal coupling our action is strictly speaking in the \textit{Jordan frame}, as opposed to the \textit{Einstein frame}, where all the couplings to gravity are canonically normalised.
In the Jordan frame, the Einstein equations and thus the backround evolution of the Universe are modified by the non-minimal coupling to \(\phi\).
In appendix \secref{sec:Jordan_Einstein_frames} we study this modification by analysing the dynamics in both frames.
We find that for the purposes of this paper, we can safely study our system in the Jordan frame without taking into account backreaction effects, knowing that they remain small.
The reason is that we are dealing with a positive coupling \(\xi\) and field values that will always remain way below the Planck scale.
The same logic also allows us to safely identify the value of the Planck mass in both frames and fix \(M\simeq\mpl\).
With this we can now proceed to analysing the evolution of the field.

\subsection{Classical evolution of the homogeneous field}
Let us start with the evolution of the homogeneous field. 
Its equation of motion is
\begin{equation}\label{eq:classical_eom}
\ddot{\phi} + 3H\dot{\phi} + \left( m^2 + \xi R \right)\phi = 0.
\end{equation}
The effect of the non-minimal coupling is that of a background-dependent mass term.
The Ricci scalar in the FRW background is given by \(R=3(1-3\omega)H^2\), where \(H\) is the Hubble parameter and \(\omega=p/\rho\) is the equation of state of the fluid that dominates the expansion.
Its value varies depending on the era according to
\begin{itemize}
\item Inflation: \(R = 12H^2,\quad (H\simeq H_\mathrm{I} = \,\mathrm{const})\).
\item Radiation domination: \(R \simeq 0,\quad (H\sim 1/(2t))\).
\item Matter domination: \(R = 3H^2,\quad (H\sim 2/(3t))\).
\end{itemize}
We expect that the positive contribution to the mass coming from the Ricci scalar will dominate during inflation and force the field to evolve towards smaller values.
During radiation domination, however, the dependence on \(R\) vanishes and the field will be constant because of the Hubble friction until we reach the time when \(3H\simeq m\) and damped oscillations start.
By the time of matter-radiation equality, the \(R\)-term will be negligible compared to the mass, as \(H_{\mathrm{eq}}\simeq 2\cdot10^{-28}\) eV, and observations require \(m\gg 10^{-28}\) eV.

The interesting regime is the inflationary one.
If we assume that \(\sqrt{12\xi}H_\mathrm{I} \gg m\), the solution to the equation of motion is
\begin{equation}\label{eq:homogeneous_eom}
\phi(t) = \frac{\phi_0}{2} \left( \me^{-\frac{\alpha_+}{2}H_\mathrm{I}t} + \me^{-\frac{\alpha_-}{2}H_\mathrm{I}t} \right),
\end{equation}
where 
\begin{equation}
\alpha_\pm = 3\pm\sqrt{9-48\xi}.
\end{equation}
For \(0<\xi<3/16\), the solution decays exponentially and if \(\xi>3/16\), there will be oscillations with an exponentially damped amplitude.
In both cases, we can relate the amplitude of the field at the end of inflation \(\phi_\mathrm{E}\) to its preinflationary value by
\begin{equation}
\phi_\mathrm{E} \simeq \phi_0\ \me^{-\frac{\mathrm{Re}(\alpha_-)}{2}N},
\end{equation}
where \(N\) is the number of e-folds of inflation.

This shows that for $\xi>0$ any possible initial homogeneous value for the field will be quickly damped away by inflation (unless \(\xi\) is very small), and the contribution to the energy density of the Universe of the homogeneous mode is totally negligible. 
We are interested in a situation where the dark matter arises from fluctuations and not from the misalignment mechanism, which justifies why in the following we will only consider values $\xi>0$. 
We also find that in most of the region where the fluctuations are sufficient to produce the full dark matter density and isocurvature constraints are avoided, $\xi$ will be sufficiently large such that $\mathrm{Re}(\alpha_-)N/2\gg 1$ and contributions from initially non-vanishing field values are small.

\subsection{Evolution of fluctuations and dark matter density}\label{flucevolution}

The goal of this section is twofold.
First, we obtain the present-day field power spectrum by tracking the evolution of the primordial one through the different eras of expansion of the Universe.
Then, we compute the energy density in the field \(\phi\) and compare it with the observed dark matter abundance.
With this, we find that in this scenario the non-minimally coupled field can account for all of the dark matter in our Universe.

The central quantity to describe fluctuations is their power spectrum.
To set our conventions, throughout this work we will use the following definition for the power spectrum of a homogeneously and isotropically distributed random field \(\Psi\):
\begin{equation}\label{eq:power_spectrum_definition}
\braket{\Psi^\star(\mathbf{k},t)\ \Psi(\mathbf{k^\prime},t)} \equiv (2\gpi)^3 \delta^{(3)}(\mathbf{k}-\mathbf{k^\prime})\frac{2\gpi^2}{k^3}\mathcal{P}_\Psi(k,t).
\end{equation}
Note that the power spectrum \(\mathcal{P}_\Psi\) can only depend on the magnitude of the momentum, \(k=|\mathbf{k}|\).
With this definition, the variance of a field is given by, 
\begin{equation}
\braket{\Psi^2} = \int\mathop{\d(\log k)}\  \mathcal{P}_\Psi(k,t).
\end{equation}
If \(\Psi\) is Gaussian-distributed, then it is completely determined by its variance (assuming it has zero mean).

For modes that are already classical, the effect of the evolution on the power spectrum of the field at any point in time is given by\footnote{At late times $\phi(k,t)$ will be oscillating quickly. In most of our calculations we will not keep track of these oscillations (e.g. by using a different symbol) but instead consider the envelope of the oscillations for which we find simple power law behaviors. When calculating the energy density from the power spectrum this is actually correct because it accounts for both potential and kinetic energy contributions according to the virial theorem, as we will explain in ~\secref{sec:energy_density_regularisation}.},
\begin{equation}
\mathcal{P}_\phi (k,t) = \mathcal{P}_\phi (k,t_0) \left( \frac{\phi (\mathrm{k}, t)}{\phi (\mathrm{k}, t_0)} \right)^2,
\end{equation}
where \(\phi (\mathrm{k}, t)\) is a solution of the classical equation of motion
\begin{equation}\label{eq:classical_eom_k}
\left(\partial^2_t + 3H\partial_t + \frac{k^2}{a^2} + m^2 + \xi R \right)\phi(k,t) = 0,
\end{equation}
with initial condition \(\phi (\mathrm{k}, t_0)\) at \(t_0\).
But even in the quantum regime the power spectrum is given in terms of the mode functions $f_{\mathbf{k}}$ of the creation and annihilation operators that evolve according to the classical equations of motion,
\begin{equation}
\mathcal{P}_{\phi} (k,\tau) = \frac{1}{a^2(\tau)} \frac{k^3}{2\gpi^2} \left| f_k(\tau) \right|^2.
\end{equation}
We will discuss this in more detail in \secref{sec:quantum_computation}. 
Here it suffices to say that the most important input from the quantum calculation is the proper choice of initial conditions for the evolution. 
In addition, the quantum formalism also provides us with a way to treat the UV-divergences that arise due to the small-scale fluctuations and obtain a finite expression for the energy density of the field.

\bigskip

\begin{figure}[h!]
\centering
\includegraphics[width=0.99\textwidth,height=\textheight,keepaspectratio]{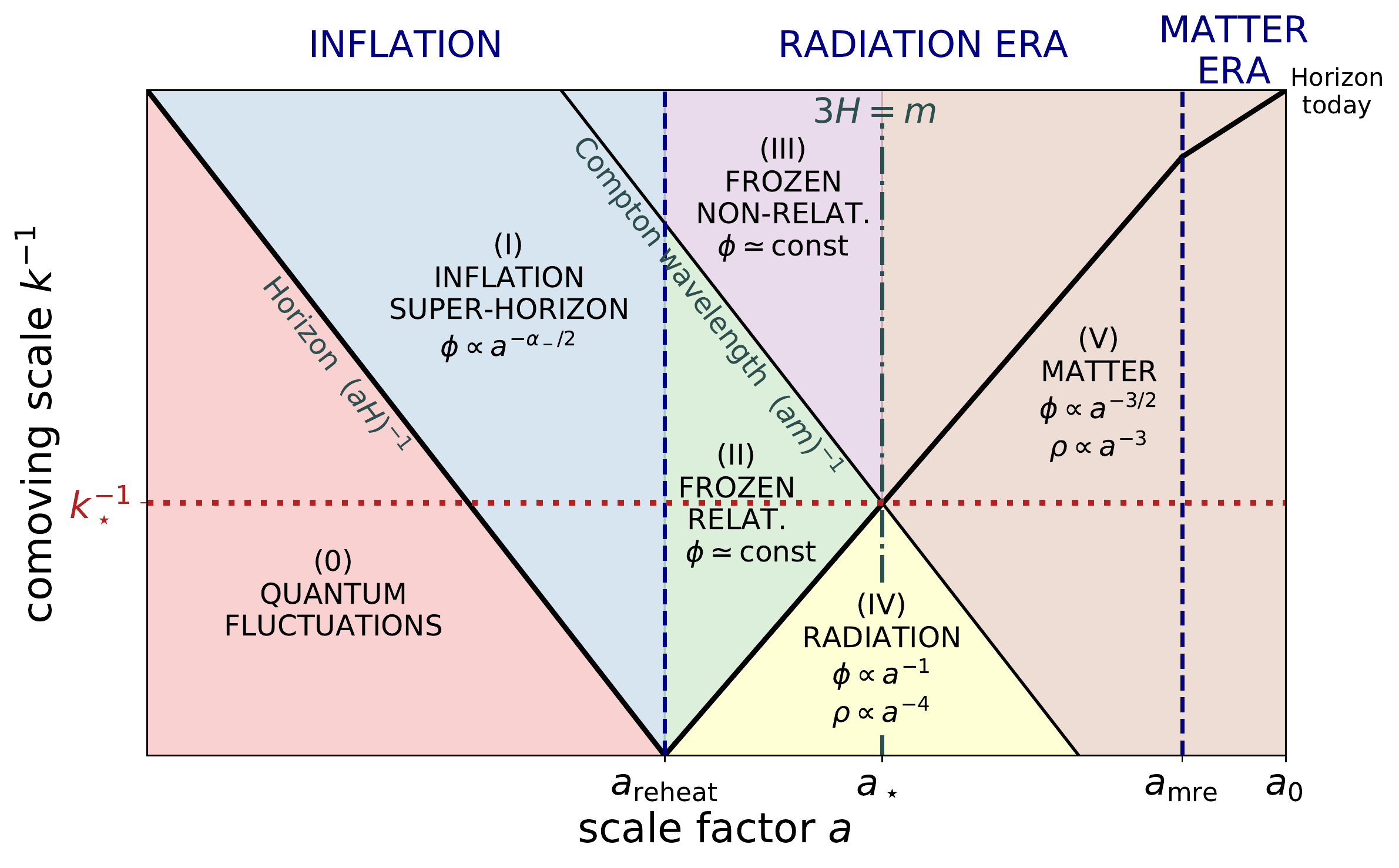}
\caption{Cosmological evolution of the amplitude  of the Fourier modes as a function of their comoving wavelength \(k^{-1}\).
The various regions for which we have analytically solved the evolution are shaded in different colours and labelled by roman numerals.
The thick black line labelled ``Horizon" represents the comoving Hubble radius \((aH)^{-1}\), which shrinks during inflation and grows during the radiation and matter eras.
The thin black line depicts the Compton wavelength of the vector \((am)^{-1}\), which dictates when a mode becomes non-relativistic.
A mode of fixed comoving wavenumber \(k\) evolves from left to right in a horizontal line, starting deep within the horizon during inflation as a quantum fluctuation and eventually behaving as a massive matter-like scalar at present times.
In between, the field amplitude behaves as depicted.
Modes around \(k_\star\), defined as the one that becomes non-relativistic at the same time that it leaves the horizon, are suppressed the least and dominate the energy density.}
\label{fig:master_plot}
\end{figure}

For now let us plow ahead and understand the relevant phenomenological features. 
In what follows, we solve the classical equation of motion \eqref{eq:classical_eom_k} analytically in the different regimes of the history of the Universe, similar to the procedure used in~\cite{graham_vector_2016}.
The relevant features of this evolution can be understood by looking at \figref{fig:master_plot}, which summarises the evolution of the different Fourier modes through those regimes, represented by different colours in that figure.

\begin{enumerate}[label=(\Roman*),start=0]
\zeroitem \textbf{Sub-horizon inflationary regime and horizon exit}.
This is the regime where quantum fluctuations are amplified by the effect of the time-evolving gravitational background until they transition into a regime when they can be treated classically. This is a crucial process that provides the appropriate initial conditions and which we will study in more detail in the next section \secref{sec:quantum_computation}. 
The bottom line is that after horizon exit the field can be described in terms of classical Fourier modes, each of them evolving independently according to \eqref{eq:classical_eom_k}.
The initial condition at horizon exit that we obtain in \secref{sec:quantum_computation} is that \(\phi_0(\mathbf{k})\) is Gaussian-distributed with power spectrum
\begin{equation}\label{eq:power_spectrum_horizon_exit_alpha}
\mathcal{P}_{\phi_0}(k) = \left( \frac{H_\mathrm{I}}{2\gpi} \right)^2 F(\alpha_-); \qquad F(\alpha_-) \equiv \frac{2^{2-\alpha_-}}{\gpi} \Gamma^2\left( \frac{3-\alpha_-}{2} \right).
\end{equation}
This is readily obtained from \eqref{eq:power_spectrum_horizon_exit}.
We see that the primordial power spectrum right at horizon exit is scale invariant, only differing from the one of a minimally coupled, massless field by the \(\orderof{(1)}\) factor \(F(\alpha_-)\).
\item \textbf{Super-horizon inflationary regime} \(H\simeq H_\mathrm{I} \gg k/a,\ m\). 
We can drop both the \(k/a\) and the \(m\) terms from the equation of motion \eqref{eq:classical_eom_k}, which then reads
\begin{equation}
\left( \partial_t^2 + 3H_\mathrm{I}\partial_t + 12\xi H_\mathrm{I}^2 \right)\phi \simeq 0\quad \longleftrightarrow\quad \left( \partial_a^2 + \frac{4}{a}\partial_a + \frac{12\xi}{a^2} \right)\phi \simeq 0.
\end{equation}
The solution is
\begin{equation}\label{eq:region_(I)}
\phi = C_+ a^{-\frac{\alpha_+}{2}} + C_- a^{-\frac{\alpha_-}{2}},
\end{equation}
with, as in \eqref{eq:homogeneous_eom}, \(\alpha_\pm = 3\pm \sqrt{9-48\xi}\). Its real part ranges from \(\mathrm{Re}(\alpha_-)=0\) for \(\xi=0\) to \(\mathrm{Re}(\alpha_-)=3\) for \(\xi\geq 3/16\).
For \(\xi>0\), the dominant term is the second one, so after a short time the solution will approach \(\phi\propto a^{-\alpha_-/2}\).
In the overlapping region, this is in agreement\footnote{Note that \(\alpha_-\) used here is related to \(\nu\) used in \secref{sec:quantum_computation} by  \(\alpha_-\simeq3-2\nu\).} with the quantum result \eqref{eq:primordial_power_spectrum}.
This result makes explicit that, after horizon exit, the effect of the non-minimal coupling quickly becomes important and suppresses the amplitude of the modes.
\item \textbf{Super-horizon radiation era relativistic regime} \(H \gg k/a \gg m\).
This regime is similar to the previous one except that during radiation domination, the Ricci scalar vanishes and the Hubble parameter is not constant.
We are then left with the equation of motion
\begin{equation}
\left( \partial_t^2 + 3H\partial_t \right)\phi \simeq 0\quad \longleftrightarrow\quad \left( \partial_a^2 + \frac{2}{a}\partial_a \right)\phi \simeq 0,
\end{equation}
which has the solution
\begin{equation}
\phi = C_1 + C_2 a^{-1}.
\end{equation}
The second term dies off and we find that in this regime \(\phi\simeq\mathrm{const}\) and the field is frozen.
\item \textbf{Super-horizon radiation era non-relativistic regime} \(H \gg m \gg k/a\).
The equation of motion is the same as the one in (II), so we conclude that the field is also frozen, \(\phi\simeq\mathrm{const}\).
\item \textbf{Sub-horizon radiation era relativistic regime} \( k/a \gg H,\ m\).
The modes behave in this region as pure radiation.
The equation of motion in physical and conformal time reads
\begin{equation}
\left( \partial_t^2 + 3H\partial_t + \frac{k^2}{a^2} \right)\phi \simeq 0\quad \longleftrightarrow\quad \left( \partial_\tau^2 + 2aH\partial_\tau + k^2 \right)\phi \simeq 0.
\end{equation}
Treating the Hubble friction term as a small perturbation, one can find the approximate solution
\begin{equation}\label{eq:region_(IV)}
\phi \simeq \frac{1}{a}\left( C_1\me^{ik\tau} + C_2\me^{-ik\tau} \right).
\end{equation}
The amplitude of the field decreases as \(\phi\propto a^{-1}\), while the energy density
\begin{equation}
\rho \simeq \frac{k^2}{a^2}\phi^2 \propto a^{-4},
\end{equation}
as expected for a relativistic fluid.
\item \textbf{Non-relativistic massive regime} \(m \gg H,\ k/a \).
The equation of motion is now
\begin{equation}
\left( \partial_t^2 + 3H\partial_t + m^2 \right)\phi \simeq 0.
\end{equation}
In this region, the mass has overcome the Hubble friction and the field performs the well-known oscillations (cf., e.g., \autocite{arias_wispy_2012}) by which it acquires a pressureless matter-like equation of state \(\omega\simeq 0\).
We can find an analytical expression for the solution of the equation of motion using the WKB approximation,
\begin{equation}
\phi\simeq a^{-\frac{3}{2}} \left( C_1 \me^{imt} + C_2 \me^{-imt} \right).
\end{equation}
The energy density in this regime redshifts as matter,
\begin{equation}
\rho \sim \half m^2\phi^2 \propto a^{-3}. 
\end{equation}
\end{enumerate}
All these solutions are summarised in \figref{fig:master_plot}.
The results obtained here for the non-minimally coupled scalar present similarities with the ones obtained in \autocite{graham_vector_2016} for the vector, but differ in several key aspects.
Firstly, the behaviour of the modes within the horizon and at horizon exit is in our scenario different to the standard massless case, due to the presence of the non-minimal coupling to gravity.
This requires us to study this quantum regime in detail, which we do in \secref{sec:quantum_computation}, together with a careful treatment of the divergencies that we encounter in the energy density of the UV modes.
Secondly, the suppression of the amplitude of the large scale modes happens in our case during inflation, when the gravitation-induced effective mass is large.
This can be seen in \eqref{eq:region_(I)} and is also calculated in section \secref{sec:quantum_computation}, where we perform the full quantum computation all the way through inflation and through the inflation-radiation era transition.
In contrast, for the vector case this suppression happens later in the evolution after the modes become non-relativistic, and is due to the particularities of the equation of motion for vectors, as discussed in \autocite{graham_vector_2016}.
Regarding the total energy density, we find that due to the evolution in the super-horizon regime we have effectively an extra suppression compared to the vector case, resulting in a preference for somewhat higher masses. 

In general, at the boundaries of each of these regimes (0)-(V), the parameters entering the equation change gradually and smoothly so that it is a good approximation to just ``glue" together the solutions that we have obtained in each of the regions.
However, there is one transition that needs special attention: reheating.
Between inflation and radiation domination, the Ricci scalar drops from \(R\simeq 12H^2\) to \(R\simeq 0\).
The exact dynamics of reheating will dictate how exactly this transition happens.
In this work we assume that reheating proceeds quickly enough that we can take it to be instantaneous and thus there is a sudden drop in the Ricci scalar.
Assuming that this transition is fast allows us to neglect any potential change in the field amplitude during reheating.
We look at this in detail in \secref{sec:quantum_computation} and confirm that this approximation is justified within our assumptions.

More generally, gravitational particle production can occur during preheating and reheating. This has been studied extensively in~\autocite{ford_gravitational_1987, chung_superheavy_1998, chung_gravitational_2001, ema_gravitational_2015, ema_gravitational_2016, markkanen_dark_2017, ema_production_2018}.
In particular, the authors of~\autocite{markkanen_dark_2017} study dark matter production through tachyonic resonance during reheating in the same non-minimally coupled setup as ours.
This effect depends on the precise model of inflation assumed, but occurs quite generically after inflation if the inflaton coherently oscillates around the minimum of its potential.
In that case, the Ricci scalar \(R\) will oscillate about zero and the \(\xi \phi^2 R\) term will periodically give rise to a tachyonic mass for the dark matter particle.
As shown in detail in~\autocite{dufaux_preheating_2006}, this can result in significant particle creation.
Following the same strategy as~\autocite{markkanen_dark_2017} but for small $\xi$, the amount of dark matter generated in this way in our setup during reheating is
\begin{eqnarray}\label{eq:reheating}
\frac{\Omega_{\rm reh}}{\Omega_{\rm DM}} \!\!&\lesssim&\!\! \xi^{3/8} \left( \frac{m}{10\,\mathrm{GeV}} \right) \left( \frac{T_{\rm reh}}{10^{15}\,\mathrm{GeV}} \right)^3 \left( \frac{H_{\rm osc}}{H_{\rm reh}} \right)^{3/4} \mathrm{exp}\left({0.73\,\frac{\sigma_{\rm osc}}{\mpl}}\right)
\\[0.5cm]\nonumber
\!\!&\ll&\!\! 1\qquad\qquad\qquad\qquad\qquad\qquad\qquad\qquad\qquad{\rm for}\quad \xi\lesssim 1,\,\,m\lesssim10\,{\rm GeV}.
\end{eqnarray}
Here, \(T_{\rm reh}\) and \(H_{\rm reh}\) denote the temperature and Hubble parameter at the end of reheating, while \(H_{\rm osc}\) and \(\sigma_{\rm osc}\) are the Hubble scale and the field value of the inflaton when it starts oscillating around the minimum of its potential (for most models, \(\sigma_{\rm osc}\lesssim \mpl\)).
Eq.~\eqref{eq:reheating} shows that this contribution to the energy density is always small if \(\xi\lesssim 1\) or \(m\lesssim 10\,\mathrm{GeV}\), both of which are satisfied in the parameter region of interest for this work, as can be seen in~\figref{fig:dark_matter}.
It is thus justified to neglect the contribution to the particle production from preheating and reheating.

\subsubsection{Field power spectrum}

With this information we are able to compute the power spectrum of the field amplitude at late times.
By late times, we mean a time when all the relevant modes are non-relativistic, so that they are behaving like matter.
In order not to spoil the correct procedure of structure formation, all modes contributing significantly to the energy density need to be non-relativistic before matter-radiation equality.
We have checked that this is indeed the case in this scenario.
The power spectra obtained in what follows are valid for all the modes that have entered region (V) (cf brown region in \figref{fig:master_plot}) and are applicable from matter-radiation equality onwards.

A simple inspection of \figref{fig:master_plot} can give a good qualitative idea of the field power spectrum.
All modes start with the same amplitude given by \eqref{eq:power_spectrum_horizon_exit_alpha} when they exit the horizon during inflation.
Then, they evolve differently depending on how much time they spend in each of the regions (I) - (V).
By visual inspection, it is easy to understand that \(k_\star\) is the mode whose amplitude is the least suppressed (at least as long as \(\alpha_-\) is not too large, we will shortly quantify this).
On the one hand, modes with large comoving wavelength (above \(k_\star^{-1}\) in \figref{fig:master_plot}) leave the horizon earlier during inflation and spend extra time being suppressed in that era.
On the other hand, modes with comoving wavelength shorter than \(k_\star^{-1}\) experience a stronger suppression while behaving like radiation in region (IV).

\begin{figure}[t]
\centering
\includegraphics[width=0.8\textwidth,height=\textheight,keepaspectratio]{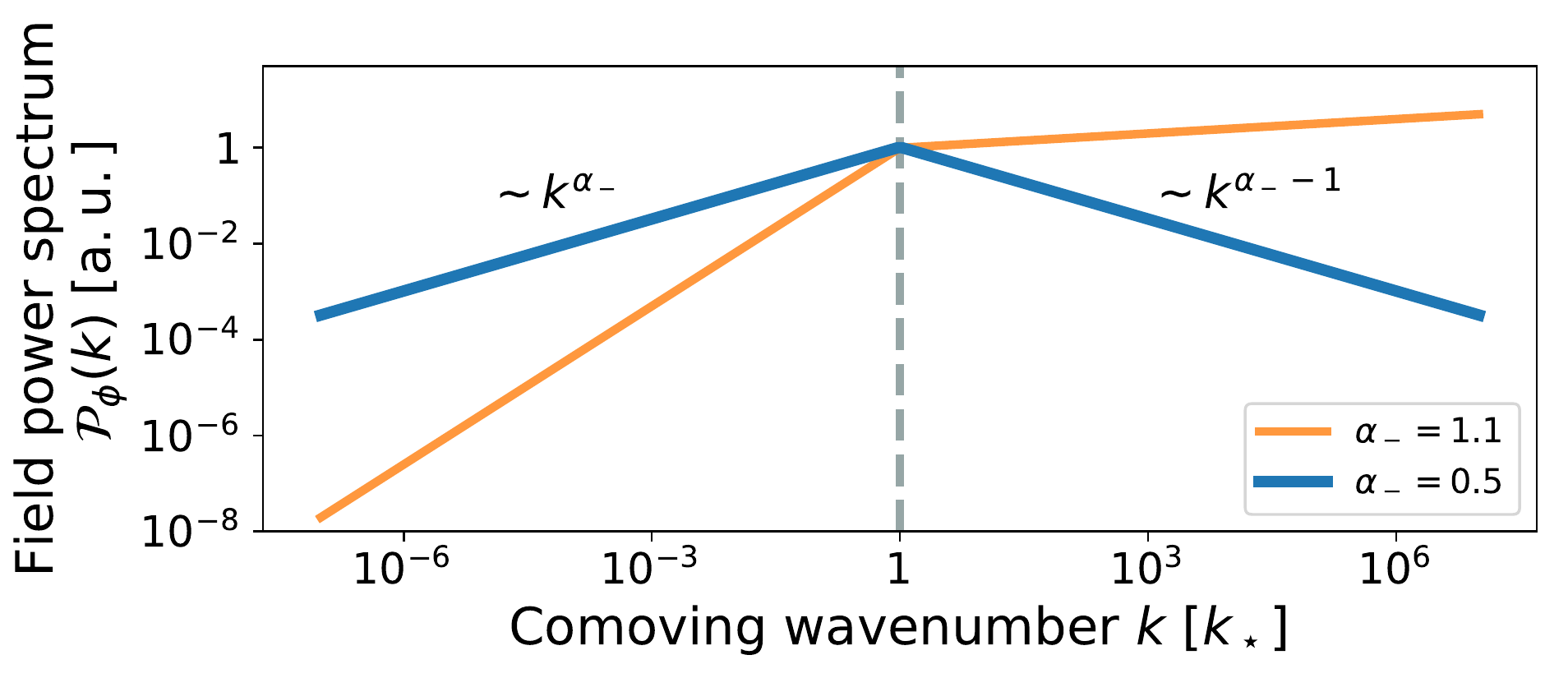}
\caption{Field power spectrum generated by inflationary fluctuations of a scalar field non-minimally coupled to gravity (normalized to 1 at \(k_{\star}\)).
The solid blue line shows the behaviour of the power spectrum for small non-minimal couplings \(\alpha_-<1\) and has a peaked structure at the critical scale \(k_\star\) (marked by the grey dashed line), which is a very small scale in cosmological terms.
The solid orange line corresponds to stronger couplings \(\alpha_->1\) and is dominated by the power at very small scales (UV modes).}
\label{fig:field_power_spectrum}
\end{figure}

This behaviour is confirmed by our calculations (see Appendix \secref{sec:power_spectra} for the details), resulting in the field power spectrum presented in \figref{fig:field_power_spectrum}.
At scales larger than the critical wavelength \(k_\star^{-1}\), the power spectrum is suppressed by \((k/k_\star)^{\alpha_-}\), while at smaller scales its amplitude scales as \((k/k_\star)^{1-\alpha_-}\).
Thus, we find that \(\alpha_-=1\) (i.e \(\xi=5/48\)) is the critical value for which the power spectrum changes from being dominated by the \(k_\star\) mode to being dominated by ultraviolet modes.
We restrict our study to the regime where \(\alpha_-<1\) featuring a peaked power spectrum.
Though potentially interesting, the scenario with \(\alpha_->1\) is dominated by the UV modes and would require a careful treatment of the divergencies with a full regularisation scheme.

These arguments agree with the result obtained from the quantum computation in \secref{sec:quantum_computation}.
Indeed, for adequate values of the non-minimal coupling, the field power spectrum in \figref{fig:quantum_power_spectrum} already exhibits a peaked structure at some intermediate scale, while it is suppressed at both very large and very small scales.
Note that the power spectrum of \figref{fig:quantum_power_spectrum} corresponds to much earlier times than those of \figref{fig:field_power_spectrum}.
This explains why the slope in both figures does not match: \figref{fig:quantum_power_spectrum} still has to be evolved through the full radiation era.

The mode with critical comoving wavenumber \(k_\star\) reenters the horizon during the radiation era precisely when it becomes non-relativistic, as can be seen in \figref{fig:master_plot}.
Phenomenologically, \(k_\star^{-1}\) is a very small scale, much below the cosmological scales that play the key roles in structure formation or can be observed with the CMB.
It is defined by \(k_\star = a_\star m\), where \(a_\star\) is the scale factor at the time when the Hubble parameter equals one third of the mass of the scalar field, \(3H(a_\star)=m\).
The comoving scale associated with this critical wavenumber can be given as a function of the mass as
\begin{equation}\label{eq:k_star}
\frac{1}{k_\star} \simeq 4.1\cdot 10^{7}\ \mathrm{km}\ \sqrt{\frac{\mathrm{eV}}{m}}.
\end{equation}
The last thing we need to completely determine the field power spectrum is to compute its overall dimensionful normalisation (note that \(\mathcal{P}_\phi\) has mass dimension 2).
To do this, we obtain the power at the critical wavenumber \(k_\star\).
We refer the reader to Appendix \secref{sec:power_spectra} for the details of the calculation.
Here we just give the final result
\begin{equation}\label{eq:power_spectrum_normalisation}
\mathcal{P}_\phi (t,k_\star) \simeq \mathcal{P}_{\phi_0}(k_\star)\ H_\mathrm{I}^{- \frac{\alpha_-}{2}}\ m^{-\half(3-\alpha_-)}\ H_{\mathrm{eq}}^{\frac{3}{2}}\ \left( \frac{a(t)}{a_{\mathrm{eq}}} \right)^{-3}.
\end{equation}
Here, \(H_\mathrm{I}\) is the Hubble scale of inflation, while the subscript ``eq" refers to quantities evaluated at the time of matter-radiation equality.
For the sake of clarity, in this expression we have removed some \(\orderof (1)\) factors.
The full version of the result is given in \eqref{eq:powerexpr}.
The primordial power spectrum at horizon exit \(\mathcal{P}_{\phi_0}\) is given in \eqref{eq:power_spectrum_horizon_exit_alpha}.
As was stressed before, this expression is only applicable at late times for modes that are already redshifting as matter.

\subsubsection{Dark matter abundance}
\begin{figure}[t]
\centering
\includegraphics[width=0.8\textwidth,height=\textheight,keepaspectratio]{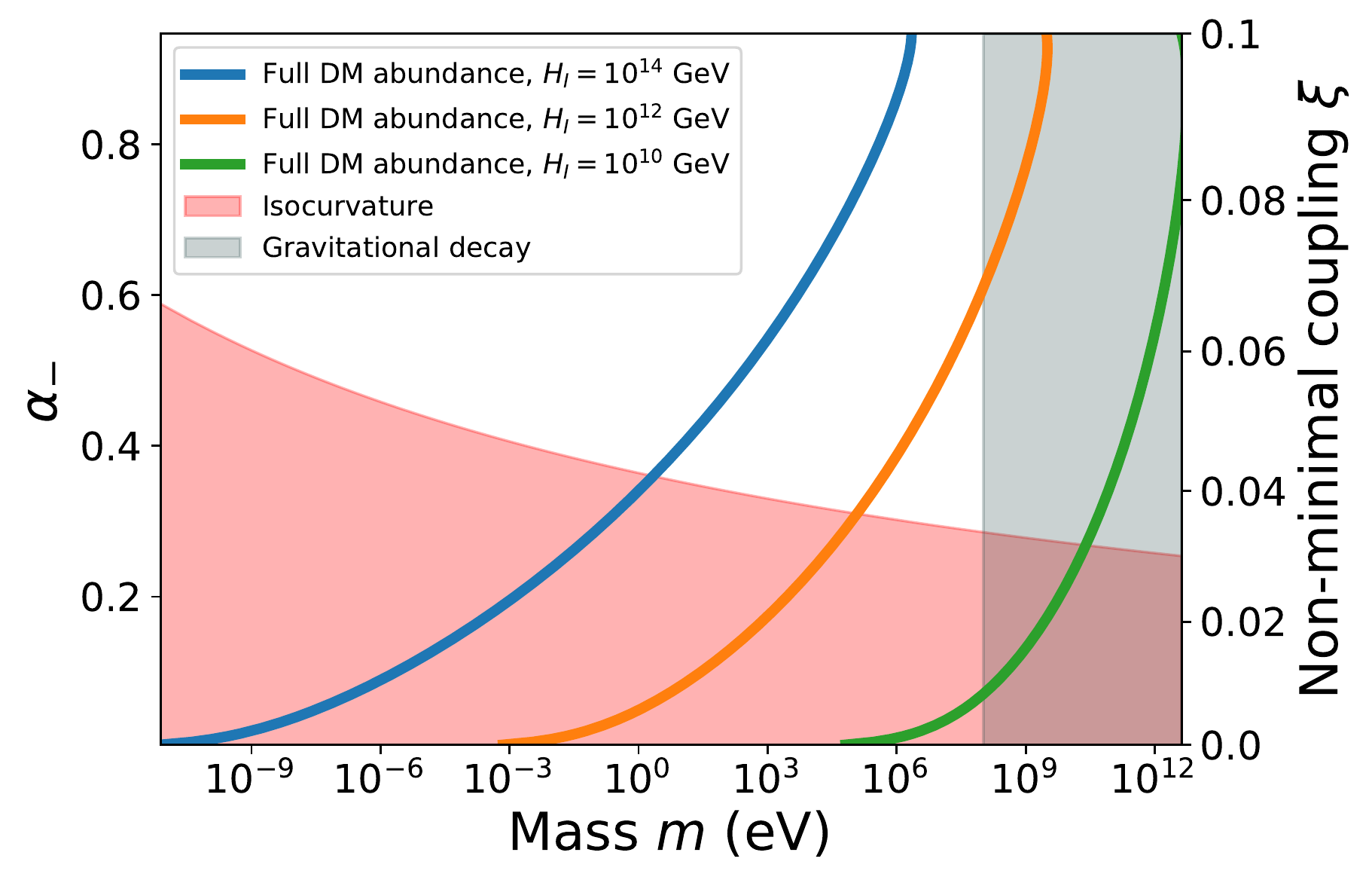}
\caption{Parameter space of our scenario, composed of the mass of the scalar field and the strength of its non-minimal coupling to gravity.
Additionally, our predictions depend on the Hubble scale of inflation \(H_\mathrm{I}\).
The lines represent the combination of masses and couplings that give the correct dark matter abundance, plotted for various values of \(H_\mathrm{I}\). 
The upper limits on the scale of inflation make it impossible to generate enough dark matter to the left of the blue line. 
The region shaded in red is excluded because our setup generates too much isocurvature power at the scales probed by the Planck satellite (note that this limit assumes that all the dark matter originates from the scalar field fluctuations and effectively only applies to the right of the blue line). 
The grey-shaded area is disfavoured if we assume that the \(\mathbb{Z}_2\) symmetry \(\phi\rightarrow -\phi\) of the field is broken by quantum gravity effects, inducing gravity-mediated decays of the dark matter particles, as discussed in \secref{sec:conclusions}.}
\label{fig:dark_matter}
\end{figure}

Once we have the power spectrum of the field, we move on to compute the dark matter abundance.
The full expression for the energy density of the field is given in next section's Eq.~\eqref{eq:energy_density_phi}.
At late times when $H$ is negligible and all relevant modes are non relativistic we have\footnote{We note that the \(\mathcal{P}_\phi (k,t)\) at this point denotes the envelope of the oscillating power spectrum as explained in \secref{sec:energy_density_regularisation}.},
\begin{equation}\label{eq:energy_density_late_times}
\rho (t) = \int \d(\log k)\ \half  m^2 \mathcal{P}_\phi (k,t).
\end{equation}
This result is independent of the regularisation scheme that has to be applied to cure the Hubble scale-dependent divergences.
The integral can be easily carried out analytically for any value of \(\alpha_-\).
Here we will focus on the regime \(\alpha_-<1\), and we refer to Appendix \secref{sec:power_spectra} for the computation and the results for other choices of \(\alpha_-\).
We can now compare the energy density in the non-minimally coupled scalar field with the observed dark matter abundance, finding the result
\begin{equation}\label{eq:dark_matter_abundance}
\frac{\Omega_\phi}{\Omega_{\mathrm{DM}}} \simeq \frac{3^{\frac{1}{2}(1+\alpha_-)}}{2^{\frac{11}{4}}\gpi^2}\, \frac{F(\alpha_-)}{\alpha_-\left( 1- \alpha_- \right)} \frac{\left[\mathcal{F}(T_\star)\right]^{1+\frac{1}{3}\alpha_-}}{\left[\mathcal{F}(T_\mathrm{r})\right]^{\frac{1}{3}\alpha_-}} \frac{1}{\mpl^2}\, H_{\mathrm{eq}}^{-\frac{1}{2}}\ H_\mathrm{I}^{\half (4-\alpha_-)}\ m^{\half (\alpha_-+1)}.
\end{equation}
Here, $\mathcal{F}(T)$ is a function encoding the dependence on the number of degrees of freedom, defined in~\cite{arias_wispy_2012} and Eq.~\eqref{ffunction}. It is typically of order $1$.
The total abundance depends, as expected, on the Hubble scale of inflation, the scalar mass and the non-minimal coupling \(\xi\) through the parameter \(\alpha_-=3-\sqrt{9-48\xi}\).
If we want to explain the observed dark matter abundance, we need to require \(\Omega_\phi=\Omega_{\mathrm{DM}}\).
This requirement, together with a choice of the Hubble scale of inflation, gives a one to one correspondence between the mass of the scalar particle and its non-minimal coupling to gravity, as can be seen in \figref{fig:dark_matter}.

\subsection{Density power spectrum and isocurvature fluctuations}\label{sec:isocurvature}
In this subsection we will compute the power spectrum of the energy density fluctuations of the non-minimally coupled field \(\phi\).
It is clear that these fluctuations are produced independently of the inflaton ones, and we thus expect both to be uncorrelated.
As the inflaton is responsible for setting the curvature (or \textit{adiabatic}) perturbations that affect the metric, the ones generated in the non-minimally coupled scalar will be of purely entropic (or \textit{isocurvature}) nature.

This is not a problem while the non-minimally coupled field remains a subdominant contribution to the energy density of the Universe.
Things change when the field's contribution to the energy budget becomes appreciable, which is of course the case in the scenario where \(\phi\) makes up the dark matter.
Then, isocurvature perturbations leave their imprint in observables like the cosmic microwave background (CMB) and the structure formation dynamics.
The lack of observation of such an imprint in the CMB by the Planck mission sets the strongest constraints on the amplitude of isocurvature perturbations \autocite{planck_collaboration_planck_2018}.
The Planck collaboration gives the name ``cold dark matter isocurvature (CDI)" to the scenario where the dark matter fluid has a contribution of isocurvature fluctuations.
In this setting, the power in isocurvature fluctuations is observed to be much smaller than the power in adiabatic ones.

This may seem like a nail in the coffin of our setup, as all the power in the non-minimally coupled field energy density is generated as isocurvature.
However, the key point is that the bulk of the energy density in our scenario is stored in the fluctuations at the very small scales around \(k_\star^{-1}\) given in \eqref{eq:k_star} to be $k_\star^{-1}\sim 1.3\times 10^{-12}\,{\rm Mpc}\sqrt{{\rm eV}/m}$.
This scale is orders of magnitude below the typical scales that the CMB and other cosmological observables probe, which typically lie in the \(k^{-1} \sim10-1000\ \mathrm{Mpc}\) range.
At those very large scales, the energy density of \(\phi\) acquires the adiabatic fluctuations of the metric, as argued in \secref{sec:adiabatic_fluctuations}.
The isocurvature power at the scales probed by Planck can in this way be suppressed below the observed limits provided. 

\begin{figure}[t]
\centering
\includegraphics[width=0.8\textwidth,height=\textheight,keepaspectratio]{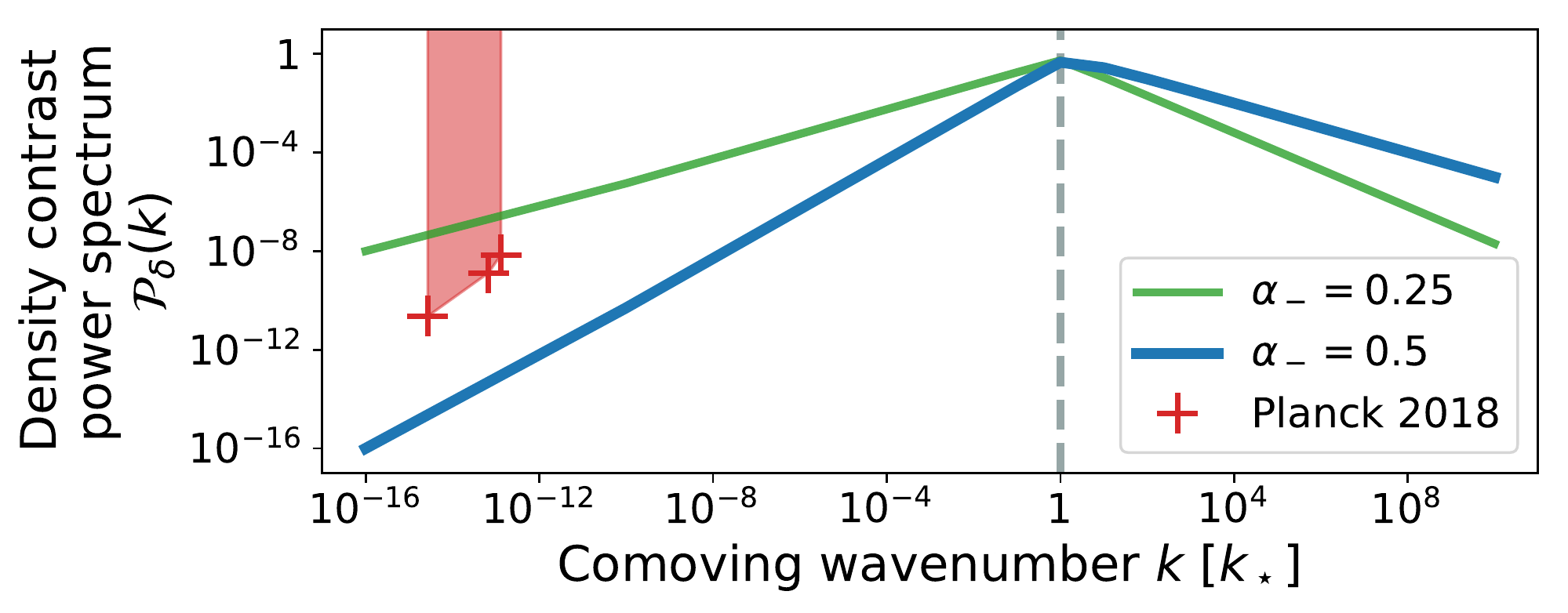}
\caption{Power spectrum of the density contrast \(\delta\) of the non-minimally coupled scalar field \(\phi\).
The general shape is valid for all values of the coupling \(\alpha_-<1\).
For this plot we have chosen \(m=1\) eV, \(\alpha_-=0.5\) (solid blue line) and \(\alpha_-=0.25\) (solid green line).
As the field power spectrum, the density contrast power spectrum is peaked at the scale \(k_\star^{-1}\) (highlighted by the grey dashed line), which is orders of magnitude smaller than the ones probed by the CMB.
The red crosses correspond to the upper limits on isocurvature perturbations (CDI) set by the Planck satellite \autocite{planck_collaboration_planck_2018}.
Note that for a large enough \(\alpha_-\), the amplitude of the isocurvature perturbations at the CMB scales is strongly suppressed. 
}
\label{fig:density_power_spectrum}
\end{figure}

Let us then compute the power spectrum of the energy overdensities of the field \(\phi\).
We define the density contrast field, that describes the deviations from the mean energy density, by
\begin{equation}
\rho(\mathbf{x}) \equiv \braket{\rho} \left( 1+\delta(\mathbf{x}) \right).
\end{equation}
Its Fourier transform can be computed as
\begin{equation}
\delta(\mathbf{k}) = \frac{1}{\braket{\phi^2}}\int \frac{\mathop{\d^3\mathbf{q}}}{(2\gpi)^3}\phi(\mathbf{q})\phi(\mathbf{k}-\mathbf{q}).
\end{equation}
The mean of the density contrast of course vanishes, and the two point function is given by
\begin{equation}
\braket{\delta^\star(\mathbf{k})\delta(\mathbf{k^\prime})} = 8\gpi^4 \delta^{(3)}(\mathbf{k}-\mathbf{k^\prime}) \frac{1}{\braket{\phi^2}^2}\int \frac{\mathop{\d^3\mathbf{q}}}{(2\gpi)^3}\ \frac{1}{q^3} \frac{1}{|\mathbf{k}-\mathbf{q}|^3} \mathcal{P}_\phi(q) \mathcal{P}_\phi(|\mathbf{k}-\mathbf{q}|).
\end{equation}
The details of this and the following calculations can be found in Appendix \secref{sec:power_spectra} (carrying out steps similar to~\cite{graham_vector_2016}).
From this we can easily extract the power spectrum of \(\delta\), which can be expressed as a function of the power spectrum of the field as
\begin{equation}\label{eq:density_power_spectrum}
\mathcal{P}_\delta(k) = \frac{k^2}{\braket{\phi^2}^2}\int_{|k-q|<p<k+q}\d q\ \d p\ \frac{1}{q^2} \frac{1}{p^2} \mathcal{P}_\phi(q) \mathcal{P}_\phi(p).
\end{equation}
This integral can be carried out numerically resulting in the density contrast power spectrum presented in \figref{fig:density_power_spectrum}.
In the parameter region of interest \(\alpha_-<1\), the density power spectrum has the same peaked structure as the field power spectrum of \figref{fig:field_power_spectrum}, and thus the energy density is dominated by overdensities of size \(\sim k_\star^{-1}\), which corresponds to the cosmologically small scale given in \eqref{eq:k_star}.
At scales smaller than \(k_\star^{-1}\) the power spectrum is suppressed by \((k/k_\star)^{(\alpha_--1)}\), whereas for large scales the suppression goes with \((k/k_\star)^{2\alpha_-}\) and is thus stronger than for the field power spectrum.

We can now compute the amplitude of isocurvature perturbations at large scales and compare them with the Planck observations to constrain our scenario.
At scales larger than the critical scale \(k_\star^{-1}\), we have
\begin{equation}
\mathcal{P}_\delta (k) \simeq \left(\frac{k}{k_\star}\right)^{2\alpha_-} \mathcal{P}_\delta (k_\star),
\end{equation}
where the power at the critical scale \(\mathcal{P}_\delta (k_\star)\) also depends on \(\alpha_-\) and can be computed using \eqref{eq:density_power_spectrum}.
We have checked that for the values that interest us, \(\mathcal{P}_\delta (k_\star) = \orderof (1)\) is an order one number.

The absence of observed isocurvature fluctuations in the CMB puts an upper limit on their amplitude. 
Planck gives constraints on the CDI scenario at three different scales \(k_i\) (see table \ref{tab:Planck_isocurvature_limits}), as upper limits for the \textit{primordial isocurvature fraction} defined as
\begin{equation}
\beta_{\mathrm{iso}}(k_i) = \frac{\mathcal{P_I}(k_i)}{\mathcal{P_I}(k_i) + \mathcal{P_R}(k_i)},
\end{equation}
where \(\mathcal{P_I}\) and \(\mathcal{P_R}\) are the isocurvature and adiabatic power spectra, respectively.
Using the Planck normalisation for the curvature power spectrum, \(\mathcal{P_R} = 2.1(9)\cdot 10^{-9}\), approximately valid for all the scales of interest \(k_i\), we can translate the bound on \(\beta_{\mathrm{iso}}\) into a bound on \(\mathcal{P_I}\).
The values are given in table \ref{tab:Planck_isocurvature_limits}.
We can see that the bounds get weaker at smaller scales.

\begin{table}[t]
\centering
\begin{tabular}{| c | c | c |}
  \hline
  \(k_i\) & \(\beta_{\mathrm{iso}}(k_i)\) & \(\mathcal{P_I}(k_i)\) \\
  \hline
  \(k_{\mathrm{low}} = 0.002\ \mathrm{Mpc}^{-1}\) & \(0.011\) & \(2.3\cdot 10^{-11}\) \\
  \(k_{\mathrm{mid}} = 0.05\ \mathrm{Mpc}^{-1}\) & \(0.38\) & \(1.3\cdot 10^{-9}\) \\
  \(k_{\mathrm{high}} = 0.1\ \mathrm{Mpc}^{-1}\) & \(0.77\) & \(7.0\cdot 10^{-9}\) \\
  \hline
\end{tabular}
\caption{Planck 2018 limits on the cold dark matter isocurvature (CDI) scenario \autocite{planck_collaboration_planck_2018}. More precisely we take the ``axion II'' scenario of that paper which is appropriate for our case.}
\label{tab:Planck_isocurvature_limits}
\end{table}

Putting everything together, we can obtain the following bound on the non-minimal coupling to gravity:
\begin{equation}
\alpha_- > \half \log\left( \frac{\beta_{\mathrm{iso}}(k_i)}{1-\beta_{\mathrm{iso}}(k_i)} \frac{\mathcal{P_R}(k_i)}{\mathcal{P}_\delta (k_\star)} \right) \left[ \log \left( \frac{k_i}{7.5\cdot10^{11}\ \mathrm{Mpc}^{-1}} \sqrt{\frac{\mathrm{eV}}{m}} \right) \right]^{-1},
\end{equation}
which assumes that the scalar makes up all of the dark matter in the Universe.
This rules out the region of parameter space shaded in red in \figref{fig:dark_matter}.
As expected, a minimum value of the non-minimal coupling \(\xi\) is required to suppress the isocurvature perturbations at large scales.

\subsection{Acquisition of adiabatic fluctuations}\label{sec:adiabatic_fluctuations}
In order for the scenario to agree with cosmological observations, it is not enough that the isocurvature perturbations are sufficiently small at large scales.
It is also necessary that the dark matter presents the observed adiabatic spectrum of fluctuations at those scales.
Here we will reason that this is indeed the case.

Because of the strongly peaked structure of the power spectrum, the energy density of the scalar field is stored in clumps of typical comoving size \(k_\star^{-1}\).
As we have argued, this is a scale orders of magnitude smaller than the ones relevant for the CMB and structure formation.
At those large scales the inhomogeneities of size \(k_\star^{-1}\) are indistinguishable from a true continuum, and the energy density of the field looks very homogeneous.
The situation is similar to other scenarios where dark matter has substructure at very small scales like, for instance, primordial black holes \autocite{carr_primordial_2016} or axion miniclusters \autocite{hogan_axion_1988,Kolb:1993zz}.

At cosmological scales, the energy density of the field is primordially homogeneous (except for the very suppressed component of isocurvature fluctuations).
After inflation ends, the fluctuations of the inflaton are imprinted in the metric in the form of curvature perturbations.
At that time, the energy density of the scalar field is subdominant and will thus follow the metric fluctuations sourced by the dominant species, which at those early times are the relativistic decay products of the inflaton.
This is the reason why the scalar field acquires the adiabatic perturbations at large scales, even if it did not have them originally.
At the same time, the small isocurvature component at large scales does not experience such a growth and becomes subdominant to the adiabatic mode.
This effect is analogous to other dark matter scenarios like axions \autocite{turner_formation_1983, axenides_development_1983} and other coherently oscillating scalar fields \autocite{iliesiu_constraining_2014, cembranos_cosmological_2016}, vectors \autocite{nelson_dark_2011-1,arias_wispy_2012,graham_vector_2016}, and further non-thermal mechanisms of dark matter production.

%% file: Quantum_computation.tex

\section{Generation of fluctuations in a curved background}\label{sec:quantum_computation}
So far we have argued mostly based on the classical equations of motion, which is essentially correct for the description of the evolution of the field after its quantum fluctuations have been amplified during inflation. 
Importantly, however, the initial conditions for this evolution are set by the quantum nature of the particle production due to the time-dependent gravitational background during the inflationary epoch.
Moreover, the quantum description also leads to UV divergencies that need to be treated. 
In this section we will discuss both of these aspects to complement and justify some of the claims made in \secref{sec:dark_matter}.

\subsection{Quantum behaviour during inflation}
We use the tools of quantum field theory in classical curved spacetimes, following \autocite{linde_particle_2005, mukhanov_introduction_2007} to obtain the power spectrum of the field arising from inflationary quantum fluctuations.
The action for the non-minimally coupled scalar field is given in \eqref{eq:action}.
We specify to a FRW background but this time using conformal time \(\tau\), defined by \(\d t = a\, \d \tau\).
The discussion is simplified by using the field redefinition
\begin{equation}
f(\tau, \mathbf{x}) = a(\tau) \phi(\tau, \mathbf{x}).
\end{equation}
Focusing only on the terms in \eqref{eq:action} with dependence on the field \(\phi\), the action becomes
\begin{equation}
S = \int \mathop{\d\tau}\mathop{\d^3\mathbf{x}} \half \left( f^{\prime 2} - |\mathbf{\nabla} f|^2 - \mu^2 f^2 \right),
\end{equation}
where the prime denotes a derivative with respect to conformal time and we have defined the effective mass
\begin{equation}
\begin{aligned}
\mu^2 &\equiv a^2 m^2 - \left( \frac{1}{6} - \xi \right) a^2 R \\
&= a^2 m^2 - \left( 1 - 6\xi \right) \frac{a^{\prime\prime}}{a} \\
&= a^2 m^2 - \left( 1 - 6\xi \right) \left( \mathcal{H}^{\prime} + \mathcal{H}^2 \right).
\end{aligned}
\end{equation}
Here, \(\mathcal{H}\) denotes the conformal Hubble parameter (we specify to inflation later in \eqref{eq:scale_factor_inflation}),
\begin{equation}
\mathcal{H}=\frac{a'}{a}.
\end{equation}
The equation of motion for the field is
\begin{equation}
f^{\prime\prime} - \mathbf{\nabla}^2 f + \mu^2 f =0.
\end{equation}
We now quantise the theory.
In the canonical quantisation formalism, we promote both \(f\) and its conjugate momentum \(\pi\) to operators and impose equal-time commutation relations
\begin{equation}\label{eq:field_momentum_operators}
\begin{aligned}
\left[ \hat f (\tau,\mathbf{x}),\ \hat \pi (\tau,\mathbf{x^\prime}) \right] &= i\delta^{(3)}(\mathbf{x}-\mathbf{x^\prime}), \\
\left[ \hat f (\tau,\mathbf{x}),\ \hat f (\tau,\mathbf{x^\prime}) \right] &= 0, \\
\left[ \hat\pi (\tau,\mathbf{x}),\ \hat\pi (\tau,\mathbf{x^\prime}) \right] &= 0.
\end{aligned}
\end{equation}
The conjugate momentum for our field is
\begin{equation}
\pi(\mathbf{x}) = \frac{\delta \mathcal{L}}{\delta f^{\prime}(\mathbf{x})} = f^{\prime} (\mathbf{x}).
\end{equation}
We can then perform expansions of both operators in Fourier space,
\begin{equation}
\begin{aligned}
\hat{f}(\tau,\mathbf{x}) &= \int \frac{d^3k}{(2\gpi)^3}\left( \hat a (\mathbf{k}) f_k(\tau) \me^{i\mathbf{k}\cdot\mathbf{x}} + \hat a^\dagger (\mathbf{k}) f^\star_k(\tau) \me^{-i\mathbf{k}\cdot\mathbf{x}} \right) \\
\hat{\pi}(\tau,\mathbf{x}) &= \int \frac{d^3k}{(2\gpi)^3}\left( \hat a (\mathbf{k}) f^\prime_k(\tau) \me^{i\mathbf{k}\cdot\mathbf{x}} + \hat a^\dagger (\mathbf{k}) f^{\prime\star}_k(\tau) \me^{-i\mathbf{k}\cdot\mathbf{x}} \right).
\end{aligned}
\end{equation}
Here, \(\hat a^\dagger,\ \hat a\) are creation and annihilation operators.
Because of isotropy, the mode functions \(f_k\) only depend on \(k = |\mathbf{k}|\), and \(f\) being real further requires that \(f_k=f_{-k}\).
The creation and annihilation operators are time-independent and satisfy commutation relations
\begin{equation}
\begin{aligned}
\left[ \hat a (\mathbf{k}),\ \hat a^\dagger (\mathbf{k^\prime}) \right] &= (2\gpi)^3\delta^{(3)}(\mathbf{k}-\mathbf{k^\prime}), \\
\left[ \hat a (\mathbf{k}),\ \hat a (\mathbf{k^\prime}) \right] &= 0, \\
\left[ \hat a^\dagger (\mathbf{k}),\ \hat a^\dagger (\mathbf{k^\prime}) \right] &= 0.
\end{aligned}
\end{equation}
This expansion and our convention for the creation and annihilation operators put a constraint on the mode functions in the form of a Wronskian normalisation,
\begin{equation}\label{eq:wronskian}
f_k(\tau)f^{\prime\star}_k(\tau) - f^\star_k(\tau)f^\prime_k(\tau) = i.
\end{equation}
As \(f_k\) and \(f^\star_k\) are linearly independent (this can be seen from the fact that the Wronskian is nonzero), the mode functions satisfy the same equation of motion as the classical fields.
Namely,
\begin{equation}\label{eq:eom_mode_function}
f_k^{\prime\prime} + \omega_k^2\,f_k = 0,\qquad \omega_k^2 (\tau) = k^2 + \mu^2(\tau).
\end{equation}
This is the Mukhanov-Sasaki \autocite{sasaki_gauge-invariant_1983,kodama_cosmological_1984,mukhanov_quantum_1988} equation for \(f\).
The time dependence of the curved background induces a time-dependent frequency for the field.
As a consequence, the vacuum state can have a non-trivial evolution into an excited state, which can be interpreted as a particle production phenomenon.
Indeed, we will see that the field will acquire fluctuations and its modes \(f_k\) will be populated.

Let us now specify to the case of an inflationary background.
For simplicity, we will assume exact de Sitter expansion with a parametrisation such that the scale factor is given by
\begin{equation}\label{eq:scale_factor_inflation}
a(\tau) = \frac{-1}{H_\mathrm{I}\,\tau},
\end{equation}
where we take the scale of inflation \(H_\mathrm{I}\) to be constant and conformal time runs from \(\tau \rightarrow -\infty\) at the beginning of inflation to \( \tau \rightarrow - \tau_r\) at the end. 
Note that, as is customary, in our conventions conformal time is negative during inflation, ensuring that the scale factor is positive, as it should be.

The other ingredient we need to solve the quantum evolution of the field operator is the initial condition for the mode functions, which amounts to choosing a suitable initial vacuum state for our theory.
For general time-dependent backgrounds this procedure can be ambiguous, but for an inflationary background there is a preferred choice.
The modes start their evolution deep inside the horizon when \(\tau \rightarrow -\infty\).
In this limit, \(k/a \gg H_\mathrm{I}\), which means that the effects of curvature become negligible.
Thus, at very early times all the modes start being approximately massless.
In this high frequency limit quantisation in de Sitter and Minkowski spaces should be equal.
The most common choice of vacuum is to select the positive frequency mode \(f_k \propto \me^{-ik\tau}\) as the minimal excitation state.
This choice, together with the Wronskian normalisation condition \eqref{eq:wronskian} sets our initial condition to be
\begin{equation}\label{eq:initial_condition_mode_function}
f_k(\tau\rightarrow -\infty) = \frac{1}{\sqrt{2k}} e^{-ik\tau}.
\end{equation}
In exact de Sitter space, the equation of motion \eqref{eq:eom_mode_function} for the mode functions together with the initial condition \eqref{eq:initial_condition_mode_function} has the following solution,
\begin{equation}
\label{eq:mode_function_inflation}
f_k(\tau) = \me^{ i\frac{\gpi}{4}(2\nu+1)}\, \frac{1}{\sqrt{2k}}\,\sqrt{ \frac{\gpi}{2}(-k\tau)}\, H_\nu^{(1)}(-k\tau), \qquad \nu^2 = \frac{9}{4} - 12\xi - \frac{m^2}{H^2}.
\end{equation}
This solution applies as long as \(\nu\notin \mathbb{N}\) and agrees with the result in \autocite{riotto_inflation_2002}.
For the range of values on which we focus in this work, the mass is much smaller than the Hubble scale of inflation, so that we can safely take \(\nu^2 \simeq \frac{9}{4} - 12\xi\).

We can now compute the quantum statistics of the field operator.
As expected, the expectation value of the field vanishes, \(\braket{\hat f} = 0\).
However, the variance of the field receives non-zero quantum fluctuations
\begin{equation}
\braket{|\hat f |^2} = \int\frac{\mathop{\d^3 \mathbf{k}}}{(2\gpi)^3} \left| f_k(\tau) \right|^2.
\end{equation}
This variance is commonly expressed as a power spectrum for the field.

With the definition~\eqref{eq:power_spectrum_definition}, we can compute the power spectrum of the original field \(\phi\) during inflation, and we obtain
\begin{equation}\label{eq:power_spectrum_during_inflation}
\begin{aligned}
\mathcal{P}_{\phi} (k,\tau) &= \frac{1}{a^2(\tau)} \frac{k^3}{2\gpi^2} \left| f_k(\tau) \right|^2 \\
&= \left(\frac{H_\mathrm{I}}{2\gpi}\right)^2 \left(-k\tau\right)^3 \frac{\gpi}{2} \left| H_\nu^{(1)}\left( -k\tau \right) \right|^2.
\end{aligned}
\end{equation}

Before we can correctly interpret our results, we have to deal with one usual problem that arises in quantum field theory in curved space-time: the power spectrum \eqref{eq:power_spectrum_during_inflation} is UV divergent.
Indeed, the variance of the field is dominated by the very small scales, where the power spectrum grows as \(k^2\).
However, we expect that modes that are deep inside the horizon are very close to the vacuum state and thus should not dominate the fluctuations.
In order to solve this issue, we should look carefully at the energy density of the field and find a suitable regularisation scheme.

\subsection{Energy density and regularisation of the power spectrum}\label{sec:energy_density_regularisation}
As we have a non-minimal coupling to the metric, we have to be careful when deriving the energy density of the field.
We start by computing the full stress-energy tensor of \(\phi\), which is given by
\begin{equation}
\begin{aligned}
T_{\mu\nu}^\phi &= \frac{-2}{\sqrt{-g}} \frac{\delta \left( \mathcal{L}_\phi\sqrt{-g} \right)}{{\delta{g^{\mu\nu}}}} \\
&= \xi \phi^2 G_{\mu\nu} + \partial_\mu\phi\partial_\nu\phi - \half \partial_\alpha\phi\partial^\alpha\, \phi g_{\mu\nu} - \half m^2\phi^2g_{\mu\nu}.
\end{aligned}
\end{equation}
Here, \(G_{\mu\nu} = R_{\mu\nu} - 1/2\,R\,g_{\mu\nu}\) is the Einstein tensor and \(\partial_0\equiv\partial_t\) is the derivative with respect to physical time.
We define the energy density through \(\rho_\phi = T_{00}^\phi\), and thus
\begin{equation}\label{eq:energy_density_phi}
\rho_\phi = \half \left( \partial_t \phi \right)^2 + \half \frac{1}{a^2} \left| \mathbf{\nabla}\phi \right|^2 + \half m^2\phi^2 +\xi G_{00}\phi^2.
\end{equation}
Let us rewrite this result in terms of the Fourier modes of the field \(f\) and using conformal time, so that we can more easily apply our previous results.
\begin{equation}\label{eq:energy_density_divergent}
a^4 \rho_\phi = \int \frac{\mathop{\d^3\mathbf{k}}}{(2\gpi)^3}\half\left[ \left| f_k^\prime \right|^2 + \left( k^2 + a^2m^2 + \left( 1-6\xi \right)\mathcal{H}^2 \right)\left| f_k \right|^2 - \mathcal{H}\left( f_k^\prime f_k^\star + f_k^{\prime\star}f_k \right) \right].
\end{equation}
It is possible to substitute the mode function for the field during inflation \eqref{eq:mode_function_inflation} in this expression.
We focus on the short wavelength modes.
If we place an arbitrary UV cutoff scale \(\Lambda\), after expanding the Hankel function for large arguments \autocite{noauthor_dlmf:_nodate} we find
\begin{equation}\label{eq:energy_density_f}
a^4\rho_\phi = C_0\Lambda^4 + C_2\mathcal{H}^2\Lambda^2 + C_4 \mathcal{H}^4\log\Lambda + \orderof{\left( \Lambda^{-2} \right)}, 
\end{equation}
where \(C_i\) are some numerical coefficients that only depend on \(\nu\) (and \(m\)).
Each one of these divergences in the energy density can be removed by adjusting the constant term, the Planck mass and the coefficient of the \(R^2\) term in the action, respectively \autocite{birrell_quantum_1982}.
We can extract a meaningful result for the energy density and the variance of the field after removing the UV divergences in this way\footnote{Of course, this is only a tree-level statement. Including loop corrections would lead to the well-known fact that an infinite number of counterterms are needed. This is the usual result that perturbative quantisation of gravity fails due to the loss of predictivity, and not in any way a particularity of our scenario.}.
To properly do this, one would need to define a consistent renormalisation scheme including the perturbative expansion of gravity.
This task is (of course) beyond the scope of this work.
Fortunately, for the purposes of this paper, a simpler regularisation will suffice.

What is observable and relevant for the purposes of dark matter production is the energy density of the field at late times.
Deep into the radiation and matter eras, the effect of the gravitational background (the expansion) is weaker, which is made explicit by the fact that \(\mathcal{H}\rightarrow 0\) in the far future.
Thus, when computing the energy density at late times, we can safely neglect the terms that depend on \(\mathcal{H}\), which are the ones that can be affected by the regularisation scheme chosen to cure the \(\mathcal{H}^2\Lambda^2\) and \(\mathcal{H}^4\log\Lambda\) divergences.

Because of this, the one divergence that we really have to care about is the \(\Lambda^4\) one.
As is well know, this one is related to the energy of the quantum vacuum and can be regularised away by a redefinition of the cosmological constant in the action.
A computation yields \(C_0=1/(4\gpi^2)\), which corresponds to the usual \(k /2\) vacuum energy of each mode in Fourier space.
The regularisation is then achieved by subtracting this quantity for every mode at the level of the energy density in \eqref{eq:energy_density_divergent}.
However, we will approach this subtraction from a slightly different perspective.
We cure this divergence by removing the corresponding term at the level of the power spectrum\footnote{Note that this transformation leaves \(\mathcal{P}_{\partial_t\phi}\) unchanged up to factors proportional to powers of \(\mathcal{H}\), which we drop at late times.} of the field with the shift
\begin{equation}\label{eq:subtraction}
\mathcal{P}_\phi (k,t) \rightarrow \mathcal{P}_\phi (k,t) - \frac{1}{2\gpi^2} \frac{k^2}{a^2}.
\end{equation}
With this, we can safely write the regularised energy density at late times as
\begin{equation}\label{fulldensity}
\rho (t) = \int \mathop{\d(\log k)}\half \left( \mathcal{P}_{\partial_t\phi}^{\mathrm{(reg)}}(k,t) + \left( \frac{k^2}{a^2} + m^2 \right) \mathcal{P}_{\phi}^{\mathrm{(reg)}}(k,t) \right),
\end{equation}
where the regularised power spectra are obtained by performing the subtraction \eqref{eq:subtraction} and then dropping all the terms proportional to powers of \(\mathcal{H}\).
This procedure of regularising the field power spectrum is effectively equivalent to the direct subtraction of the vacuum energy, as the only observable consequence of the power spectrum of the field is precisely its contribution to the energy density.
 
The regularised power spectrum at the end of inflation is shown in \figref{fig:quantum_power_spectrum}.
The fact that we have not removed the Hubble scale-dependent divergences explains why the power stays constant at small scales.
\figref{fig:quantum_power_spectrum} also shows how the power in those UV modes decreases as \(H^2\) after the end of inflation, as expected.
This shows that we can obtain a power spectrum and an energy density which are free of divergences at late times.

After being reassured that our quantum computation leads to well-defined observables at late times, we can safely extract the required initial conditions for the classical evolution computed in \secref{sec:dark_matter}.
For modes well outside the horizon that satisfy \(-k\tau\ll H_\mathrm{I}\), we can Taylor expand the regularised version of \eqref{eq:power_spectrum_during_inflation} to obtain the approximate expression\footnote{Note that regularisation has a negligible effect in the power spectrum for superhorizon modes, as required.}
\begin{equation}\label{eq:primordial_power_spectrum}
\mathcal{P}_\phi^{\mathrm{(reg)}} (k,t) \simeq \mathcal{P}_{\phi_0}^{\mathrm{(reg)}}(k)\ \left( \frac{k}{a(t)H_\mathrm{I}} \right)^{3-2\nu},
\end{equation}
where
\begin{equation}\label{eq:power_spectrum_horizon_exit}
\mathcal{P}_{\phi_0}^{\mathrm{(reg)}}(k) = \left( \frac{H_\mathrm{I}}{2\gpi} \right)^2 \frac{2^{2\nu-1}\Gamma^2(\nu)}{\gpi}
\end{equation}
is the power spectrum at horizon exit.
This is precisely the initial condition that we use for the classical evolution of the field in \secref{sec:dark_matter}.
Moreover, note that the result \eqref{eq:primordial_power_spectrum}, valid in the region labelled (I) in \figref{fig:master_plot}, matches the classical result obtained in \eqref{eq:region_(I)} (we remind the reader that \(\alpha_- = 3-2\nu\)).

From now on, we drop the superscript (reg) and it should be understood that we are always dealing with regularised power spectra.
Of course, it should also be understood that all the quantities with which we deal in \secref{sec:dark_matter} are regularised and free of divergencies.

\bigskip
A further comment is in order. As mentioned in \secref{flucevolution} we usually do not account for the rapid oscillations of
$\phi$ and in consequence $\mathcal{P}_{\phi}(k)$, but instead only consider the envelope (for which we use the same symbol). Moreover, in \secref{sec:dark_matter} we use the simpler Eq.~\eqref{eq:energy_density_late_times} instead of Eq.~\eqref{fulldensity} to calculate the energy density, where the former only includes the last term of the latter.
At late times this follows from the virial theorem. Averaging the energy density over (at least) one oscillation we have
\begin{equation}
\langle \dot{\phi}^2\rangle=\langle (m^2+k^2)\phi^2\rangle=\frac{1}{2}(m^2+k^2)\phi_{{\rm amp}}^2,
\end{equation}
where $\phi_{\rm amp}$ gives the amplitude that is specified by the enveloping function of $\mathcal{P}_{\phi}(k)$.
Inserting this into Eq.~\eqref{fulldensity} we obtain Eq.~\eqref{eq:energy_density_late_times}.

\subsection{End of inflation and radiation domination era}\label{sec:end_of_inflation}
In exact de Sitter expansion, the solution \eqref{eq:mode_function_inflation} holds exactly.
However, inflation is only an approximate de Sitter era.

Firstly, we know that, at least during the epoch when the scales observable in the CMB leave the horizon, the Hubble parameter varies slightly with time.
We nevertheless ignore this effect in our derivation, the reason being that we are particularly interested in modes around \(k_\star\) that exit the horizon much later than the ones observable in the CMB.
There is very little observational guidance to how the Hubble parameter behaves at the epoch relevant for those modes.
Rather than extrapolating the value of the spectral index orders of magnitude away from the region where it is measured, we opt for the simplest assumption of a constant Hubble scale of inflation.
In \secref{sec:conclusions} we will nevertheless discuss how this effect could be incorporated into our results.

Secondly, inflation has to end at some point so that the Universe can be reheated.
As a first approximation, we will assume that this process happens very quickly when the inflating Universe reaches a size \(a_r\).
After that moment, we assume that the Universe enters an era where the relativistic decay products of the inflation dominate the background energy density.
In our parametrisation, inflation ends at conformal time \(-\tau_r\), with \(\tau_r = 1/(a_rH_\mathrm{I})\), and the equation of state of the Universe suddenly changes from \(\omega=-1\) to \(\omega=1/3\).

The solution \eqref{eq:mode_function_inflation} is thus only valid for \(\tau<-\tau_r\).
For \(\tau>-\tau_r\), we have to solve the mode equation \eqref{eq:eom_mode_function} for a radiation dominated expansion.
Then, the time-dependent frequency simplifies to \(\omega_k^2 = k^2 + a^2m^2\), as the Ricci scalar is zero in the radiation era.
If we assume that the modes are relativistic at that time (this assumption is justified \textit{a posteriori}), then \(\omega_k^2\simeq k^2\) becomes time-independent and we can easily solve the mode equation to obtain the outgoing vacuum mode function
\begin{equation}
v_k^{(\mathrm{out})}(\tau) = \frac{1}{\sqrt{\omega_k}} \me^{i(\tau+\tau_r)\omega_k}, \qquad \tau>-\tau_r.
\end{equation}
The incoming wave function can thus be expanded after reheating via a Bogolyugov transformation as 
\begin{equation}
\begin{aligned}
f_k(\tau) &= \alpha_k^{\star}v_k^{(\mathrm{out})}(\tau) + \beta_k^\star v_k^{\star(\mathrm{out})}(\tau) \\
&= \frac{1}{\sqrt{\omega_k}}\left( \alpha_k^\star \me^{i(\tau+\tau_r)\omega_k} + \beta_k^\star \me^{-i(\tau+\tau_r)\omega_k} \right), \qquad \tau>-\tau_r.
\end{aligned}
\end{equation}
We can extract the Bogolyugov coefficients by requiring continuity of the mode function and its first derivative at \(\tau=-\tau_r\).
The result is
\begin{equation}
\begin{aligned}
\alpha_k^\star &= \me^{i\frac{\gpi}{4}(2\nu+1)} \frac{\sqrt{2\gpi}}{4} \sqrt{k\tau_r} \left[ \left( 1-i \frac{(\nu+1/2)}{k\tau_r} \right)H_\nu^{(1)}(k\tau_r) + i H_{\nu+1}^{(1)}(k\tau_r) \right],\\
\beta_k^\star &= \me^{i\frac{\gpi}{4}(2\nu+1)} \frac{\sqrt{2\gpi}}{4} \sqrt{k\tau_r} \left[ \left( 1+i \frac{(\nu+1/2)}{k\tau_r} \right)H_\nu^{(1)}(k\tau_r) - i H_{\nu+1}^{(1)}(k\tau_r) \right].
\end{aligned}
\end{equation}
The fact that \(\left| \alpha_k \right|^2 - \left| \beta_k \right|^2 = 1\) ensures that the normalisation condition \eqref{eq:wronskian} is still satisfied.
The mode function that we have derived is valid while radiation domination lasts and as long as the modes are relativistic.

\begin{figure}[t]
\centering
\includegraphics[width=0.7\textwidth,height=\textheight,keepaspectratio]{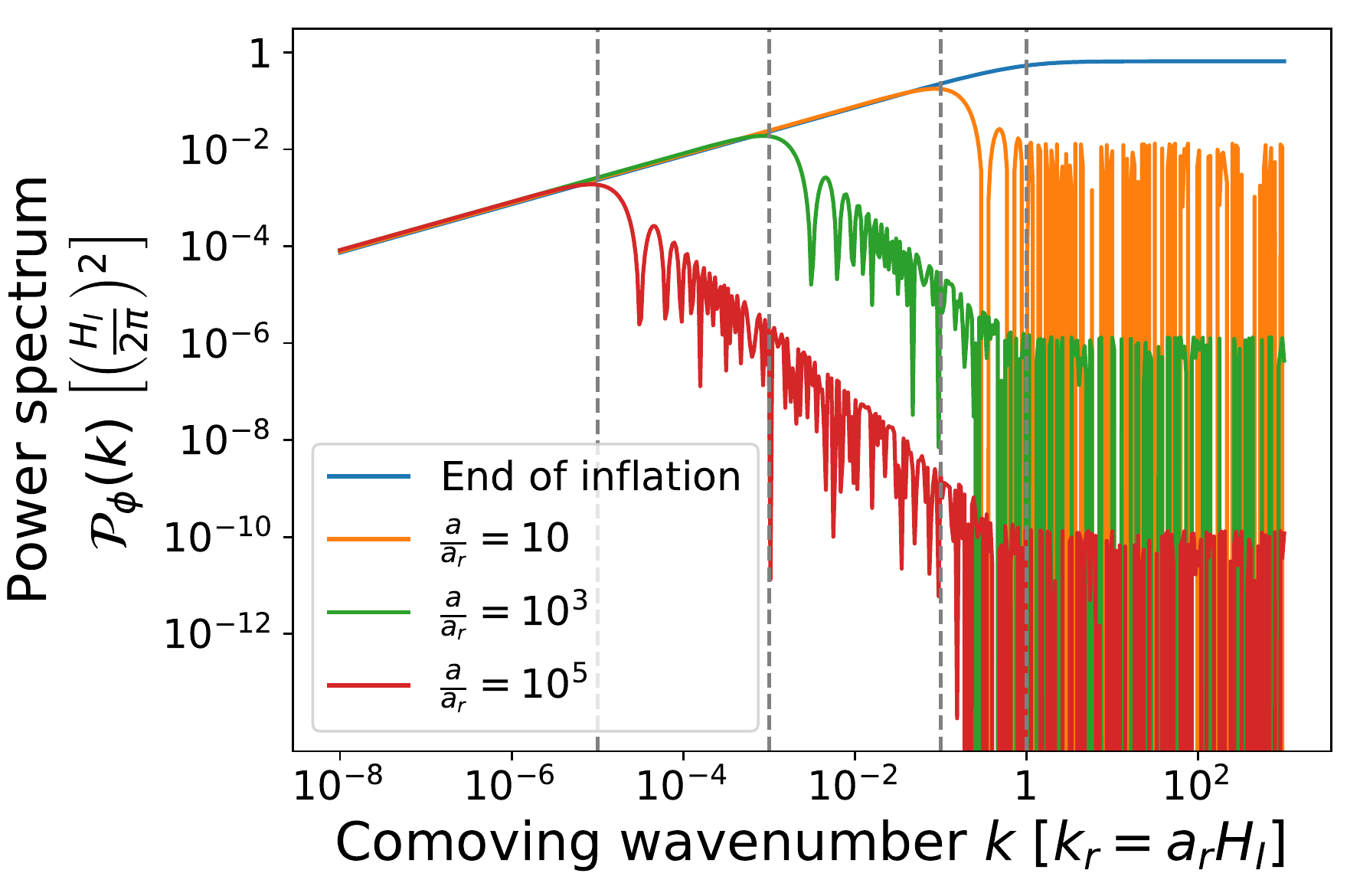}
\caption{Power spectrum of the field while the modes are still relativistic, as a function of the wavenumber normalised to the Hubble radius at the end of inflation, \(k_r^{-1}=\left(a_rH_\mathrm{I}\right)^{-1}\) (note that this normalisation is different to the one in Figures \ref{fig:field_power_spectrum} and \ref{fig:density_power_spectrum}).
The solid lines depict the power spectrum right at the end of inflation and at different moments during radiation domination.
The dashed vertical lines depict the comoving Hubble radius at the end of inflation and each later moment, respectively from right to left.
A value of \(\nu=5/4\) (equivalently $\alpha_{-}=1/2$) is chosen here, corresponding to a non-minimal coupling \(\xi \simeq 0.06 \).
}
\label{fig:quantum_power_spectrum}
\end{figure}

The power spectrum in the radiation era can then be expressed\footnote{In this expression we have already removed the contribution of the vacuum energy, as discussed in section \secref{sec:energy_density_regularisation}.} as
\begin{equation}\label{eq:power_spectrum_RD}
\mathcal{P}_\phi (k,a) = 2\left( \frac{H_\mathrm{I}}{2\gpi} \right)^2 \left( \frac{a_r}{a} \right)^2 \left( \frac{k}{a_rH_\mathrm{I}} \right)^2 \left[ 2\left| \beta_k \right|^2 + 2\mathrm{Re}\left( \alpha_k^\star\,\beta_k\,\me^{2i\frac{k}{a_rH_\mathrm{I}}\left( \frac{a}{a_r}+1 \right)} \right) \right].
\end{equation}
This power spectrum is depicted in \figref{fig:quantum_power_spectrum} at different moments of the radiation dominated era, starting right after inflation.
Wavelengths larger than \(k_r^{-1}\equiv \left(a_rH_\mathrm{I}\right)^{-1}\) exit the horizon at some point during inflation and are excited, while shorter wavelengths that stay inside the horizon do not receive large fluctuations.
The power decreases at very large scales because of the suppression arising from the effective mass during inflation due to the non-minimal coupling.
This means that right after inflation, the power spectrum is peaked at the scale \(k_r^{-1}\).
Later on, as modes reenter the horizon during radiation domination, they start to oscillate and their amplitude is damped, while superhorizon modes are still frozen.
The consequence is that the peak in the power spectrum shifts towards larger scales.
This will continue as long as all the modes involved are relativistic, which means that the final position of the peak will be at the scale which reenters the horizon at the same time as it becomes non-relativistic, which happens when \(H(a_\star)=m\).
Note that the suppression for small-scale modes in \figref{fig:quantum_power_spectrum} is steeper than in \figref{fig:field_power_spectrum}.
This is due to the fact that \figref{fig:quantum_power_spectrum} depicts the power spectrum at early times \(a\ll a_\star\), and thus the modes inside the Hubble radius are still relativistic.
After \(a_\star\), these subhorizon small-scale modes start becoming non-relativistic and the final shape \figref{fig:field_power_spectrum} is attained (this becomes clearer by looking at the evolution outlined in \figref{fig:master_plot}).

It is possible to make analytical approximations of the power spectrum \eqref{eq:power_spectrum_RD} for the modes with \(k \ll k_r\), that is, modes that were well outside the Hubble radius at the end of inflation.
We can distinguish two regimes, depending on wether the modes have already reentered the horizon during radiation domination or not.
If they have, we can average out the oscillating terms and find
\begin{equation}
\mathcal{P}_\phi (k,\tau) \simeq \left( \frac{H_\mathrm{I}}{2\gpi} \right)^2 \frac{1}{\gpi}\, 2^{2(\nu-1)} \Gamma^2(\nu) \left( \nu - 1/2 \right)^2 \left( \frac{k}{a_rH_\mathrm{I}} \right)^{1-2\nu} \left( \frac{a_r}{a(\tau)} \right)^2,
\end{equation}
which is valid if \(k\gg aH \).
In this expression, the damping of the power with the scale factor is manifest in the last factor and matches the result \eqref{eq:region_(IV)} corresponding to region (IV) in \figref{fig:master_plot}.
In the opposite limit, \(k\ll aH \), the modes are still superhorizon and an expansion of the oscillating functions for small arguments gives
\begin{equation}
\mathcal{P}_\phi (k,\tau) \simeq \left( \frac{H_\mathrm{I}}{2\gpi} \right)^2 \frac{1}{\gpi}\, 2^{2\nu-1} \Gamma^2(\nu) \left( \nu - 1/2 \right)^2 \left( \frac{k}{a_rH_\mathrm{I}} \right)^{3-2\nu},
\end{equation}
confirming that the amplitude of these long wavelength modes is constant as long as they remain superhorizon (regions (II) and (III) in \figref{fig:master_plot}).

Let us recapitulate: In this section we have seen how a light scalar field which is non-minimally coupled to gravity acquires quantum fluctuations that can grow large during inflation.
We have been careful to study the UV divergences that arise and we have extracted a regularised, physically meaningful power spectrum.
Interestingly, the existence of the non-minimal coupling suppresses the magnitude of superhorizon modes during inflation, meaning that the amplitude of the power spectrum decreases on very large scales.
We have seen that when the modes reenter the horizon after inflation, they start to oscillate and their amplitude is also damped.
These effects result in a power spectrum which is peaked at some intermediate scale.
This quantum analysis complements the classical study done in \secref{sec:dark_matter}, in particular giving the right initial conditions for the classical evolution.
It also justifies the phenomenological results in \secref{sec:dark_matter}, showing that they are firmly grounded and independent of the particular UV physics that is necessary to make sense of the full quantum theory.

%% file: Conclusions.tex

\section{Conclusions}\label{sec:conclusions}

\subsection{Summary}
In this work, we have shown that in one of the simplest extensions of general relativity, i.e. general relativity plus a non-minimally coupled scalar field, the scalar field can acquire a relic density sufficient to account for the totality of the dark matter observed in our Universe.
In our scenario, dark matter is produced during inflation by purely gravitational means independently of any couplings to other fields that could be present.
As shown in \figref{fig:dark_matter}, for small values of the non-minimal coupling constant \(\xi\lesssim 0.1\) together with masses above \(\sim 1\) eV,  high-scale inflation naturally yields the right relic abundance. For lower inflation scales and similar non-minimal couplings significantly higher masses are viable, e.g. $m\sim 10\,{\rm keV}$ for $H_{\mathrm{I}}\sim 3\cdot10^{12}\,$GeV.

This scenario does not rely on any initial conditions, and the only necessary requirement is for the field to be present during the inflationary era.
The reason for this is that the energy density is entirely produced from the quantum fluctuations of the scalar field, which are amplified during inflation due to the strong time dependence of the scale factor.
Such fluctuations are suppressed on the largest scales due to the presence of the non-minimal coupling.
In the regime that we consider, the non-minimal coupling essentially acts as an effective mass term for the field of the order of the Hubble parameter during inflation.
At a later stage of the evolution during radiation domination, the fluctuations on small scales are also suppressed due to the fact that the modes are still relativistic when they reenter the horizon and oscillate with damped amplitude.
The interplay of these two effects can be intuitively understood by looking at \figref{fig:master_plot}.
The resulting energy density power spectrum is depicted in \figref{fig:density_power_spectrum}.
The main feature is that it is peaked at some intermediate scale \(k_\star^{-1}\) given by \eqref{eq:k_star}, which means that the energy density is mainly stored in clumps of typical comoving size \(k_\star^{-1}\).
Compared to a treatment of the field as quasi-homogeneous~\cite{bertolami_scalar_2016,cosme_scale-invariant_2018} there are two implications. The first is the presence of additional small scale substructure (similar to the vector case~\cite{graham_vector_2016}). Secondly, the suppression of the relativistic higher momentum modes during the radiation dominated era leads to an overall reduction in the dark matter density for a given mass and inflation scale. In consequence, sufficient dark matter production requires somewhat higher masses.  

An important observation is that this scenario predicts that the density perturbations of the dark matter are of isocurvature type, as they are uncorrelated with the perturbations in the other fluids.
However, these fluctuations are only large at intermediate scales close to \(k_\star^{-1}\sim 4\times 10^{7}\,{\rm km}\), while their amplitude is suppressed at the very large scales relevant for cosmological observables such as the CMB (cf.~\cite{graham_vector_2016,bertolami_scalar_2016,cosme_scale-invariant_2018}).
Indeed, this suppression can be large enough for the power spectrum to fall below the stringent constraints set by the Planck satellite, as computed in \secref{sec:isocurvature} and summarised in \figref{fig:dark_matter}.

While the quantum fluctuations lead to UV divergences, we have argued that they can be absorbed into the parameters of the Lagrangian such as the cosmological constant.
We have provided a definition of the energy density that is compatible with this regularisation and thus free of divergences.
The amount of dark matter is finite.

\subsection{Discussion}
In this paper, we have focused on the basics of the mechanism and the generation of dark matter, leaving the study of the interesting phenomenology associated with the scenario for future work.
However, we would like to briefly comment on some interesting points.

As was stated in \secref{sec:end_of_inflation}, in our derivation we have assumed a constant Hubble parameter during inflation.
It is nevertheless easy to see that our results remain qualitatively unchanged if we allow for a small time variation of it.
In particular, a slowly decreasing \(H\), as a constant spectral index \(n_s<1\) hints, amounts to a constant tilt of the power spectrum, increasing towards larger scales.
This can be compensated with a slight shift towards larger values of the non-minimal coupling \(\xi\), after which all of our conclusions remain valid.
Of course, there is no reason to assume that the spectral index remains constant beyond the scales probed by the Planck satellite. However, without observational guidance that could provide a preference for any particular behaviour, this issue, though potentially very rich in terms of physical consequences, lies beyond the scope of this work.

A natural question that arises has to do with the detectability and stability of the dark matter in this scenario.
It is in principle possible to consider direct couplings of the scalar to other fields. This opens up many possibilities and would require an extensive study.
Also note that in principle the action \eqref{eq:action} allows for a \(\mathbb{Z}_2\) symmetry \(\phi\rightarrow-\phi\) which would forbid decay-inducing couplings.
We may nevertheless assume that gravity does not preserve this \(\mathbb{Z}_2\) symmetry\footnote{It is conjectured that gravity breaks all continuous global symmetries \autocite{abbott_wormholes_1989,coleman_wormholes_1990,giddings_axion-induced_1988,kallosh_gravity_1995}, but discrete symmetries can remain unbroken~\autocite{krauss_discrete_1989,kim_abelian_2013}.}.
In such a case \(\phi\) could undergo gravity-mediated decay.
The most relevant decay channels for the mass range discussed in this paper are decays to photons or light fermions.
For the decay to two photons, mediated by the operator \(\frac{\phi}{\mpl}F_{\mu\nu}F^{\mu\nu}\), the rate can be estimated as \(\Gamma\sim m^3/64\gpi \mpl^2\).
If we require that the lifetime of the dark matter is longer than the age of the Universe, we get the limit \(m\lesssim 0.1\) GeV.
The decay to two fermions can arise from the operator \(\frac{\phi}{\mpl}\bar{f}Hf\), where after electroweak symmetry breaking the Higgs is replaced by its vacuum expectation value \(v\simeq246\) GeV.
With this, we estimate the decay rate as \(\Gamma\sim N_\mathrm{c}v^2m/16\gpi \mpl^2\), assuming that the scalar is much heavier than the fermions.
This can give a much larger decay rate than the one to two photons.
In particular, any scalar heavier than \(\sim 1\) eV would be cosmologically unstable due to the decay to neutrinos.
However, the operator that we wrote down for the decay to two fermions is probably too simplistic.
For instance, it seems plausible that the operator comes with an appropriate Yukawa factor (for all we know, the fermions always couple to the Higgs this way)\footnote{This is exactly the case if the decay is induced by a linear coupling of \(\phi\) to the Ricci scalar, as was considered in \autocite{cata_dark_2016, cata_dark_2017}.}.
This would mean that the rate would go as \(m_f^2\,m\) instead of \(v^2\,m\), thus suppressing the decay to the lighter fermions.
In this case, it is easy to see that a potential decay to two photons becomes the dominant one for any scalar mass.
This stability estimate is shown in \figref{fig:dark_matter} but should only be taken as an indication and not as a strong bound, given the arguments presented above.

In contrast to the lack of precise predictions regarding the non-gravitational interactions, the mechanism has potentially interesting cosmological signatures. A first feature is the potential presence of a small component of isocurvature fluctuations also at large scales. This could be detectable\footnote{In the most recent Planck analysis~\cite{planck_collaboration_planck_2018}, the collaboration has studied a scenario (called ``axion II'') that is applicable to our isocurvature spectrum. Intriguingly the data seems to have a very slight preference for the presence of non-vanishing isocurvature fluctuations.} in the vicinity of the boundary of the red region 
in \figref{fig:dark_matter}.
The main feature is that the bulk of the energy density is contained in substructures whose characteristic comoving size, as long as they follow the Hubble flow, is \(k_\star^{-1}\).
However, as these overdensities are very large (the value of the power spectrum at the peak is close to  \(\mathcal{O}(1)\)), it is likely that they decouple from the expansion and form bound structures very early.
Indeed, the authors of \autocite{enander_axion_2017} (see also \autocite{kolb_large-amplitude_1994}) showed that this decoupling can happen around matter-radiation equality.
The characteristic size of these bound structures is then
\begin{equation}
\ell_{\mathrm{today}}\simeq \frac{1}{z_{\mathrm{eq}}}k_\star^{-1} \simeq 10^{4}\,\mathrm{km}\,\sqrt{\frac{\mathrm{eV}}{m}}.
\end{equation}
These small-scale overdensities are similar to the ones described in \autocite{graham_vector_2016} and also present some resemblance with the \textit{axion miniclusters} that are predicted to form (under some assumptions) if a cosmological population of QCD axions exists in our Universe \autocite{hogan_axion_1988,Kolb:1993zz}.
Although they are probably too small to affect the formation of large scale structures in any significant way, they could have interesting signatures in gravitational lensing~\cite{fairbairn_searching_2017-1,fairbairn_structure_2018}, astrophysical processes~\autocite{zurek_astrophysical_2007,penarrubia_fluctuations_2018} and in direct~\autocite{grabowska_detecting_2018} and indirect detection experiments, in the case where additional couplings to Standard Model fields exist.
In particular, at larger masses, indirect detection signatures, e.g. coming from an annihilation of dark matter particles, could be enhanced due to the clumpiness.

%% file: Jordan_Einstein_frames.tex

\section{Einstein and Jordan frame dynamics}\label{sec:Jordan_Einstein_frames}

The action in the form presented in \eqref{eq:action} is written in the Jordan frame where the Planck mass depends on the value of the scalar field. This is in contrast to the Einstein frame, where all the couplings to gravity are canonically normalised.
Both frames are related by a conformal transformation of the metric (cf., e.g.,~\cite{Wetterich:1987fk}).
Starting from the Jordan frame metric \(g\), we can get to the Einstein frame one \(\tilde{g}\) by means of the transformation
\begin{equation}\label{eq:conformal_trafo}
g_{\mu\nu} \rightarrow \tilde{g}_{\mu\nu} = \Omega^2(x) g_{\mu\nu}, \quad \mathrm{where} \quad \Omega^2(x) = \frac{1}{\mpl^2}\left( M^2 - \xi\phi^2 \right).
\end{equation}
Under such a transformation, we have that
\begin{equation}
\begin{aligned}
g^{\mu\nu} &\rightarrow \tilde{g}^{\mu\nu} = \Omega^{-2} g_{\mu\nu} \\
\sqrt{\left| g \right|} &\rightarrow \sqrt{\left| \tilde{g} \right|} = \Omega^4 \sqrt{\left| g \right|} \\
R &\rightarrow \tilde{R} = \Omega^{-2}R - 6\Omega^{-3}\Box\Omega .
\end{aligned}
\end{equation}
Our action then becomes
\begin{equation}
\begin{aligned}
S = \int \d^4x \sqrt{\left| \tilde{g} \right|} &\left( \half \mpl^2 \tilde{R} + \frac{1}{\Omega^4}\tilde{\mathcal{L}}_{\mathrm{back}}\left( \phi, \sigma_i \right) \right. \\
&\ \ \left. - \half \left( \frac{1}{\Omega^2}-\frac{6\xi^2\phi^2}{\mpl^2\Omega^4} \right) \tilde{g}^{\mu\nu}\partial_\mu\phi\partial_\nu\phi - \frac{1}{2\Omega^4} m^2\phi^2 \right),
\end{aligned}
\end{equation}
where \( \tilde{\mathcal{L}}_{\mathrm{back}} \) is the transformed version of the background Lagrangian accounting for all the other fields, including the Standard Model ones. It may now also depend on \(\phi\) due to the form of the transformation \eqref{eq:conformal_trafo}.

This shows that in the Einstein frame, where the coupling to gravity is minimal, non-canonical kinetic terms for the fields \(\phi\) and \(\sigma_i\) arise.
We can nevertheless canonically normalise the field \(\phi\) by performing a field redefinition to a new field \(\varphi\), given by
\begin{equation}
\frac{\d \varphi}{\d\phi} = \sqrt{\frac{1}{\Omega^2}-\frac{6\xi^2\phi^2}{\mpl^2\Omega^4}}.
\end{equation}
In terms of \(\varphi\), the action reads
\begin{equation}
S = \int \d^4x \sqrt{\left| \tilde{g} \right|} \left( \half \mpl^2 \tilde{R} + \frac{1}{\Omega^4}\tilde{\mathcal{L}}_{\mathrm{back}}\left( \varphi, \sigma_i \right) - \half \partial_\mu\varphi\partial^\mu\varphi - \frac{1}{2\Omega^4} m^2\phi^2(\varphi) \right).
\end{equation}

In general we cannot neglect the backreaction of the non-minimal coupling on the metric. 
However, if we are dealing with a not too large coupling constant \(\xi\) and subplanckian field values \(\phi \ll \mpl\), then these effects are expected to be small.
Indeed, in this regime we have that \(M\simeq \mpl\) and we can approximate
\begin{equation}\label{eq:canonical_field_transformation}
\begin{aligned}
\Omega^2 & \simeq 1 - \xi\left(\frac{\phi}{\mpl}\right)^2, \\
\varphi &= \phi \left[ 1 + \frac{3}{2} (1-6\xi)\xi\left(\frac{\phi}{\mpl}\right)^2 + \orderof\left( \left(\frac{\phi}{\mpl}\right)^4 \right) \right].
\end{aligned}
\end{equation}
This shows that backreaction effects on the metric appear at higher orders in the \(\phi/\mpl\) expansion.
To first order only the dynamics of the non-minimally coupled field \(\phi\) will be affected.
One may wonder where the effect on the field $\varphi$ is encoded.
As will become clear in the example below, this arises from the second term in \eqref{eq:canonical_field_transformation}.
Expanding the action in $\varphi$, an additional mass term for the field appears that is proportional to the density.
For sufficiently high density/Hubble scale this term can be bigger than $m^2$, significantly altering the evolution of the field.

Let us now study a specific example for the background Lagrangian  \(\mathcal{L}_{\mathrm{back}}\).
We set our metric to be a flat FLRW one,
\begin{equation}
\d s^2 = -\d t^2 + a^2(t)\left( \d x^2 + \d y^2 + \d z^2 \right).
\end{equation}
With this particular choice of geometry, the Ricci scalar reads
\begin{equation}
R = 6\left( \dot{H} + 2H^2 \right) = 3\left( 1 - 3 \omega \right) H^2,
\end{equation}
where the dot denotes a time derivative, \( H = \dot{a}/a \) is the Hubble parameter and the last equality is valid if the Universe is dominated by a perfect fluid with equation of state \(\omega = P/\rho\), being \(P\) and \(\rho\) its pressure and energy density, respectively.

\subsection{Background dominated by a cosmological constant}\label{sec:cosmological_constant_bkg}
We assume that the background is dominated by some form of vacuum energy or cosmological constant \(\Lambda\), which we parametrise as usual by
\begin{equation}
\mathcal{L}_{\mathrm{back}} = - \Lambda \mpl^2.
\end{equation}
\begin{enumerate}[label=(\alph*)]
\item \textit{Jordan frame:}

It is well known that a cosmological constant makes the Universe expand exponentially fast,
\begin{equation}
a(t) = a_0\me^{Ht} \ ,\quad \mathrm{with}\quad H^2 = \frac{\Lambda}{3} = \mathrm{const.}
\end{equation}
The curvature in this scenario is \(R = 12H^2\), and the equation of motion for \(\phi\) reads
\begin{equation}
\ddot{\phi} + 3H\dot{\phi} + \left( m^2 + 12\xi H^2 \right)\phi = 0.
\end{equation}
We see that in this frame the non-minimal coupling introduces an effective mass term for the scalar field, \(m^2_{\mathrm{eff}} = m^2 + 4\xi \Lambda\).
The solution to the equation of motion is
\begin{equation}\label{eq:field_Lambda}
\phi(t) = C_+\me^{- \frac{\alpha_+}{2}Ht} + C_-\me^{- \frac{\alpha_-}{2}Ht},
\end{equation}
where
\begin{equation}
\alpha_{\pm} = 3\pm \sqrt{9-48\xi-4\left( \frac{m}{H} \right)^2}.
\end{equation}
If \(C_-\neq 0\), the second term in \eqref{eq:field_Lambda} dominates at late times.
The condition to have a growing mode is
\begin{equation}
\alpha_- < 0 \quad \Leftrightarrow \quad \xi < - \frac{m^2}{4\Lambda}.
\end{equation}
If this is satisfied, the field value grows exponentially.
The reason is that in this range of negative couplings the effective mass \(m^2_{\mathrm{eff}}\) becomes negative and we have a runaway potential for the field \(\phi\).
In the main text we are interested in $\xi>0$ so this does not happen there.

\item \textit{Einstein frame:}

After the conformal transformation \(g_{\mu\nu}\rightarrow \hat{g}_{\mu\nu} = \Omega^2g_{\mu\nu}\) and the field redefinition \(\phi\rightarrow\varphi\) that have already been described, we get to the Einstein frame action, where the coupling to gravity is minimal and the scalar field is canonically normalised.
Next we expand the action only keep leading terms in the \(\varphi/\mpl\) expansion, as we are assuming that the field is subplanckian.
The resulting action is
\begin{equation}
\begin{aligned}
S = \int \d^4x \sqrt{\left| \tilde{g} \right|} &\left[ \half \mpl^2 \tilde{R} - \Lambda\mpl^2 - \half \partial_\mu\varphi\partial^\mu\varphi - \frac{1}{2}\left( m^2 + 4\xi\Lambda \right)\varphi^2+\orderof\left( \varphi^4\right) \right]. 
\end{aligned}
\end{equation}
Identifying $\Lambda=3H^2$ we recover the same equation of motion that we had in the Jordan frame.
Of course, if \(\varphi\) takes values close to the Planck scale our expansion in terms of \(\varphi^2/\mpl^2\) breaks down and an exact treatment is required.

\end{enumerate}
This example shows that it is reasonable to ignore backreaction effects and work directly in the Jordan frame without modifying the Einstein's equations, as long as we are only dealing with subplanckian field values. We have explicitly checked that this also holds for a background energy dominated by an additional scalar field.

%% file: Power_spectra.tex

\section{Power spectra}\label{sec:power_spectra}
\subsection{Field power spectrum}
In this appendix, we present the calculations of the power spectra in a bit more detail.
The most useful tool for this will be \figref{fig:master_plot}, where the solution of the equation of motion for the modes in each regime is summarised.
Tracking the evolution of the modes in horizontal lines from left to right in that figure, we can obtain the late-times power spectrum as
\begin{equation}
\mathcal{P}_\phi (k,t_0) = 
\left\{
\begin{aligned}
&\mathcal{P}_{\phi_0} \left( \frac{a_\mathrm{r}}{k/H_\mathrm{I}} \right)^{-\alpha_-} \left( \frac{a_0}{k_\star/m} \right)^{-3},\qquad & k<k_\star, \\
&\mathcal{P}_{\phi_0} \left( \frac{a_\mathrm{r}}{k_\star/H_\mathrm{I}} \right)^{-\alpha_-} \left( \frac{a_0}{k_\star/m} \right)^{-3},\qquad & k=k_\star, \\
&\mathcal{P}_{\phi_0} \left( \frac{a_\mathrm{r}}{k/H_\mathrm{I}} \right)^{-\alpha_-} \left( \frac{k/m}{H_\mathrm{I}a_\mathrm{r}^2/k} \right)^{-2} \left( \frac{a_0}{k/m} \right)^{-3},\qquad & k>k_\star,
\end{aligned}
\right.
\end{equation}
with \(k_\star\) as defined in \eqref{eq:k_star}.
To obtain this, we have assumed a standard cosmology with an instantaneous transition from inflation to radiation domination.
We can use the mode \(k_\star\) as a reference and write the amplitude of the other modes with respect to it,
\begin{equation}
\mathcal{P}_\phi (k,t_0) = 
\left\{
\begin{aligned}
&\mathcal{P}_{\phi}(k_\star, t_0) \cdot \left( \frac{k}{k_\star} \right)^{\alpha_-},\qquad & k\leq k_\star, \\
&\mathcal{P}_{\phi}(k_\star, t_0) \cdot \left( \frac{k}{k_\star} \right)^{\alpha_- -1},\qquad & k\geq k_\star.
\end{aligned}
\right.
\end{equation}
This confirms that, as long as \(0<\alpha_-<1\), the power spectrum is peaked at the scale \(k_\star^{-1}\), as shown in \figref{fig:field_power_spectrum}.
The next step is to evaluate the overall normalisation of the power spectrum, which we do by computing the amplitude of the mode \(k_\star\).
For that, we make use of the definition \(k_\star = a_\star m\), together with the fact that during the radiation era, \(H\propto a^{-2}\).
We find
\begin{equation}
\begin{aligned}\label{eq:powerexpr}
\mathcal{P}_{\phi}(k_\star, t_0) &= \mathcal{P}_{\phi_0} \left(\frac{H_\mathrm{I}}{m}\right)^{-\alpha}\left( \frac{a_\mathrm{r}}{a_\star} \right)^{-\alpha} \left( \frac{a_\mathrm{eq}}{a_\star} \right)^{-3} \left( \frac{a_0}{a_{\mathrm{eq}}} \right)^{-3} \\
&= \frac{3^{\frac{1}{2}(3+\alpha_-)}F(\alpha_-)}{2^{\frac{11}{4}}\gpi^2}\,\frac{\left[\mathcal{F}(T_\star)\right]^{1+\frac{1}{3}\alpha_-}}{\left[\mathcal{F}(T_\mathrm{r})\right]^{\frac{1}{3}\alpha_-}} \,H_\mathrm{I}^{\frac{1}{2}(4-\alpha_-)}\, m^{-\frac{1}{2}(3-\alpha_-)}\, H_{\mathrm{eq}}^{\frac{3}{2}} \left( \frac{a_0}{a_{\mathrm{eq}}} \right)^{-3}.
\end{aligned}
\end{equation}
Here, we have used the result of the primordial power spectrum \eqref{eq:power_spectrum_horizon_exit_alpha} at the time when the mode leaves the horizon.
We have also assumed that the comoving entropy \(s=a^3S\) is conserved since the beginning of the radiation era and until present time.
This results in the presence of the function~\autocite{arias_wispy_2012}
\begin{equation}
\label{ffunction}
\mathcal{F}(T) = \left( \frac{g_\star(T)}{3.36} \right)^{\frac{3}{4}}\cdot\left( \frac{g_{\star S}(T)}{3.91} \right)^{-1},
\end{equation}
which depends on the number of relativistic degrees of freedom and is always of order \(\orderof{(1)}\) within the temperature range of interest, assuming Standard Model degrees of freedom.

\subsection{Relic abundance}
We now move to the computation of the dark matter abundance.
We can substitute the result Eq.~\eqref{eq:powerexpr} into the expression for the energy density at late times \eqref{eq:energy_density_late_times} to find (using $\d(\log k)=\d k/k$),
\begin{equation}
\rho_\phi(t_0) = \half m^2 \mathcal{P}_\phi(k_\star, t_0) \left( \int_{\frac{k_{\mathrm{min}}}{k_\star}}^{1}\mathop{\d x}x^{\alpha_--1} + \int^{\frac{k_{\mathrm{max}}}{k_\star}}_{1}\mathop{\d x}x^{\alpha_--2} \right).
\end{equation}
Here we have explicitly kept the limits of integration.
These are essentially set by the duration of inflation.
The largest scale corresponds to the size of the Universe, \(k_{\mathrm{min}}\sim a_0 H_0\), which would be undistinguishable from a homogeneous contribution.\footnote{See~\cite{graham_stochastic_2018,guth_qcd_2018} for a way to use this as the actual source of dark matter.}
The smallest scale is determined by the Hubble radius at the end of inflation, \(k_{\mathrm{max}}\sim a_\mathrm{r}H_\mathrm{I}\), where we cut off the power spectrum.
In general, we have that \(k_{\mathrm{min}}\ll k_\star\ll k_{\mathrm{max}}\).
In the main text we consider $0<\alpha_{-}<1$ where the integral is convergent on both ends and we can safely put the lower integration limit to $0$ and the upper one to $\infty$. Let us nevertheless have a brief look at all possible cases.

It is more useful and illuminating to express the result in terms of the density parameter \({\Omega_\phi\equiv \rho_\phi / \rho_{\mathrm{crit}}}\), and compare it with the observed dark matter density.
Depending on the value of \(\alpha_-\), we have different regimes:
\begin{itemize}
\item If \(\alpha=0\Leftrightarrow \xi=0\), we recover the minimally coupled case, where the power spectrum is flat at large scales.
The energy density is dominated by the lower limit of the integral, that is, the homogeneous mode:
\begin{equation}
\frac{\Omega_\phi}{\Omega_{\mathrm{CDM}}} \simeq \frac{m^2 \mathcal{P}_\phi(k_\star, t_0)}{3H_{\mathrm{eq}}^2\mpl^2} \left( 1 - \log\left(\frac{k_{\mathrm{min}}}{k_\star}\right) \right).
\end{equation}
\item If \(0<\alpha<1\Leftrightarrow 0<\xi\lesssim 0.1\), the power spectrum is peaked at the intermediate scale \(k_\star^{-1}\).
The result is in this case independent of both lower and upper cutoffs of the power spectrum:
\begin{equation}
\frac{\Omega_\phi}{\Omega_{\mathrm{CDM}}} \simeq \frac{m^2 \mathcal{P}_\phi(k_\star, t_0)}{3H_{\mathrm{eq}}^2\mpl^2} \frac{1}{\alpha_-(1-\alpha_-)}.
\end{equation}
\item If \(\alpha=1\Leftrightarrow \xi\simeq 0.1\), the power spectrum is flat for large momenta above \(k_\star\), and the energy density is logarithmically sensitive to the UV cutoff:
\begin{equation}
\frac{\Omega_\phi}{\Omega_{\mathrm{CDM}}} \simeq \frac{m^2 \mathcal{P}_\phi(k_\star, t_0)}{3H_{\mathrm{eq}}^2\mpl^2} \left( 1 + \log\left(\frac{k_{\mathrm{max}}}{k_\star}\right) \right).
\end{equation}
\item If \(\alpha>1\Leftrightarrow \xi\gtrsim0.1\), the power spectrum is dominated by large momenta, and most of the energy is stored at the UV cutoff:
\begin{equation}
\frac{\Omega_\phi}{\Omega_{\mathrm{CDM}}} \simeq \frac{m^2 \mathcal{P}_\phi(k_\star, t_0)}{3H_{\mathrm{eq}}^2\mpl^2} \frac{1}{(\alpha_--1)} \left( \frac{k_{\mathrm{max}}}{k_\star} \right)^{\alpha_--1}.
\end{equation}
\end{itemize}
In the main text we focus on the second of these four cases, corresponding to a power spectrum peaked at intermediate scales.
The observable results being independent of the details of the beginning and the end of inflation is particularly appealing and allows us to make statements that do not rely on the specific choice of inflationary model.

\subsection{Energy density power spectrum}
In the previous section we have computed the mean energy density in the field \(\phi\), but we are also interested in the statistics of its spatial distribution.
The spatial fluctuations of the energy density with respect to the mean value can be encoded in the density contrast \(\delta\), defined by
\begin{equation}
\rho (\mathbf{x}) = \braket{\rho}\left( 1+\delta({\mathbf{x}}) \right).
\end{equation}
The density contrast is a random field, and as such we can compute its power spectrum.
In momentum space, we can write
\begin{equation}
\delta(\mathbf{k}) = \frac{1}{\braket{\phi^2}} \int \frac{\mathop{\d^3\mathbf{q}}}{(2\gpi)^3}\phi(\mathbf{q})\phi(\mathbf{k-q}).
\end{equation}
As expected, we can check that its expectation value vanishes,
\begin{equation}
\begin{aligned}
\braket{\delta(\mathbf{k})} &=  \frac{1}{\braket{\phi^2}} \int \frac{\mathop{\d^3\mathbf{q}}}{(2\gpi)^3}\braket{\phi(\mathbf{q})\phi(\mathbf{k}-\mathbf{q})} \\
&=  \frac{2\gpi^2}{\braket{\phi^2}} \int \frac{\mathop{\d^3\mathbf{q}}}{q^3} \mathcal{P}_\phi(q) \delta^{(3)}(\mathbf{k}) \\
&= 0 \qquad \mathrm{if}\ k\neq 0.
\end{aligned}
\end{equation}
To get to the second line, we have used the definition of the power spectrum and the fact that \(\phi(\mathbf{x})\) is real, and thus \(\phi^{\star}(\mathbf{k}) = \phi(-\mathbf{k})\).
The two point function does not vanish.
We can compute it using Wick's theorem in the following way
\begin{equation}
\begin{aligned}
\braket{\delta(\mathbf{k})\delta(\mathbf{k}^\prime)} &=  \frac{1}{\braket{\phi^2}^2} \int \frac{\mathop{\d^3\mathbf{q}}}{(2\gpi)^3} \frac{\mathop{\d^3\mathbf{q}'}}{(2\gpi)^3} \braket{\phi(\mathbf{q})\phi(\mathbf{k}-\mathbf{q})\phi(\mathbf{q}')\phi(\mathbf{k}'-\mathbf{q}')} \\
&= \frac{1}{\braket{\phi^2}^2} \int \frac{\mathop{\d^3\mathbf{q}}}{(2\gpi)^3} \frac{\mathop{\d^3\mathbf{q}'}}{(2\gpi)^3} 
\left[\begin{aligned} \braket{\phi(\mathbf{q})\phi(\mathbf{k}-\mathbf{q})}\braket{\phi(\mathbf{q}')\phi(\mathbf{k}'-\mathbf{q}')}\\ 
+\braket{\phi(\mathbf{q})\phi(\mathbf{q}')}\braket{\phi(\mathbf{k}-\mathbf{q})\phi(\mathbf{k}'-\mathbf{q}')}\\ 
+\braket{\phi(\mathbf{q})\phi(\mathbf{k}'-\mathbf{q}')}\braket{\phi(\mathbf{q}')\phi(\mathbf{k}-\mathbf{q})} 
\end{aligned}\right]\\
&=0 + \frac{4\gpi^2}{\braket{\phi^2}^2} \int \mathop{\d^3\mathbf{q}}\mathop{\d^3\mathbf{q}'} \frac{\delta^{(3)}(\mathbf{q}+\mathbf{q}')}{q^3} \mathcal{P}_\phi(q) \frac{\delta^{(3)}(\mathbf{q}-\mathbf{k}-\mathbf{k}'+\mathbf{q}')}{\left| \mathbf{q}-\mathbf{k} \right|^3} \mathcal{P}_\phi(\left| \mathbf{q}-\mathbf{k} \right|) \\
&\hspace{0.7cm}+ \frac{4\gpi^2}{\braket{\phi^2}^2} \int \mathop{\d^3\mathbf{q}}\mathop{\d^3\mathbf{q}'} \frac{\delta^{(3)}(\mathbf{q}+\mathbf{k}' - \mathbf{q}')}{q^3} \mathcal{P}_\phi(q) \frac{\delta^{(3)}(\mathbf{q}'+\mathbf{k}-\mathbf{q})}{q^{\prime 3}} \mathcal{P}_\phi(q') \\
&= (2\gpi)^3 \delta^{(3)}(\mathbf{k}+\mathbf{k}') \frac{2\gpi^2}{k^3}\, \frac{4\gpi^2 k^3}{\braket{\phi^2}^2} \int \frac{\mathop{\d^3\mathbf{q}}}{(2\gpi)^3} \frac{\mathcal{P}_\phi(q)}{q^3} \frac{\mathcal{P}_\phi(\left| \mathbf{q}-\mathbf{k} \right|)}{\left| \mathbf{q}-\mathbf{k} \right|^3}.
\end{aligned}
\end{equation}
From here, we can read off the power spectrum
\begin{equation}
\begin{aligned}
\mathcal{P}_\delta (k) &= \frac{k^3}{\braket{\phi^2}^2} \int \mathop{\d q}\mathop{\d \cos\theta} \frac{1}{q^3} \frac{1}{\left| \mathbf{q}-\mathbf{k} \right|^3}\mathcal{P}_\phi(q) \mathcal{P}_\phi(\left| \mathbf{q}-\mathbf{k} \right|) \\
&= \frac{k^3}{\braket{\phi^2}^2} \int_{\left| q-k \right| < p < q+k} \mathop{\d q}\mathop{\d p} \frac{1}{q^2} \frac{1}{p^2}\mathcal{P}_\phi(q) \mathcal{P}_\phi(p). 
\end{aligned}
\end{equation}
To get to the second line, we have performed a change of variables \(\cos\theta \rightarrow p\equiv \left| \mathbf{k} - \mathbf{q} \right|\).